\documentclass[superscriptaddress,aps,pre,floatfix]{revtex4-2}
\usepackage[utf8]{inputenc}
\usepackage{amsmath}
\usepackage{amssymb}
\usepackage{graphicx}
\usepackage{amsmath}
\usepackage{amssymb}
\usepackage{bm}
\usepackage{subcaption}
\usepackage{bm}
\usepackage[english]{babel}
\usepackage{graphicx}
\usepackage{xcolor}

\begin{document}

\title{Quantum Kinetic Theory of Plasmas}
\author{Gert Brodin}
\affiliation{Department of Physics, Ume{\aa} University, SE--901 87 Ume{\aa}, Sweden}
\author{Jens Zamanian}
\affiliation{Department of Physics, Ume{\aa} University, SE--901 87 Ume{\aa}, Sweden}
\date{Received: / Accepted:}

\begin{abstract}
As is well-known, for plasmas of high density and modest temperature, the classical kinetic theory needs to be extended. Such extensions can be based on the Schrödinger Hamiltonian, applying a Wigner transform of the density matrix, in which case the Vlasov equation is replaced by the celebrated Wigner-Moyal equation. Extending the treatment to more complicated models, we investigate aspects such as spin dynamics (based on the Pauli Hamiltonian), exchange effects (using the Hartree-Fock approximation), Landau quantization, and quantum relativistic theory. In the relativistic theory we first study cases where the field strength is well beyond Schwingers critical field.  Both weakly relativistic theory (gamma factors close to unity) and strongly relativistic theory are investigated, using assumptions that allow for a separation of electron and positron states. Finally, we study the Dirac-Heisenberg-Wigner (DHW) formalism, which is a fully quantum relativistic theory, allowing for field strengths of the order of the Schwinger critical field or even larger. As a result, the quantum kinetic theory is extended to cover phenomena such as Zitterbewegung and electron-positron pair creation. While the focus of this review is on the quantum kinetic models, we illustrate the theories with various applications throughout the manuscript.  
\end{abstract}

\maketitle 

\section{Introduction}
Historically, the vast majority of plasma physics has been limited to classical (non-quantum) phenomena. With important applications, such as magnetically confined fusion plasmas and space plasmas, where the density is modest and the temperature is high, the focus of plasma physics on classical models has been natural. However, in recent times, much work (see e.g. Refs. \cite{Manfredi-2019,Shukla-Eliasson-2011,Melrose-2020,Shukla-Eliasson-2011-2,Vladimirov-review}, and references therein) has been devoted to the regime of high density and a low or modest temperature, allowing for quantum mechanical phenomena to influence the dynamics. We will come back to which concrete systems that can be of interest in such a context.  

Out of the possible quantum effects, maybe the most basic one is particle dispersion (see e.g. \cite{Haas-book-chap2,ManfrediWigner-Numerics,Haas-Shukla 2008,Haas-mfl-2008}). This spread out in the wave function is captured already in the Schrödinger equation, but arguably this is one of the most important quantum mechanical mechanisms. Another basic, and important, phenomenon is the degeneracy effect \cite{Manfredi-2019,Shukla-Eliasson-2011,Melrose-2020,Shukla-Eliasson-2011-2,Vladimirov-review,Vladimirov-surf1,Vladimirov-surf2}, entering when the density becomes high enough to make the Fermi temperature significant in relation to the thermodynamic temperature. A related phenomena, also dependent on the antisymmetry of the manybody wavefunction of electrons, but considerably more complicated to model mathematically, is the exchange interaction \cite{Manfredi-exchange,Haas-exchange,Andreev-exchange,Egen-2013-exchange,Egen-2015-exchange,Egen-2015-second-exchange,Egen-2019-exchange}, whose importance probably has been undervalued in the recent plasma research literature. Upgrading from the Schrödinger Hamiltonian to the Pauli-Hamiltonian, the spin dynamics enter the picture \cite{Manfredi-spin1,Brodin2008,Zamanian 2010,Andreev-spin1,Andreev-spin2,Andreev-spin3,Andreev-spin4}, with physics such as the magnetic dipole force, spin precession, and (spin) magnetization currents. Extending the models to cover the weakly relativistic regime \cite{Asenjo-2012,Hurst-2017,Ekman-2021}, spin-orbit interaction, Thomas precession and a spin-dependent polarization current are new features of the theory. Finally, in the regime of strong fields \cite{strongfield1,strongfield2} and fully relativistic theory \cite{BB-1991}, phenomena such as strong Landau quantization \cite{landau-quant} and electron-positron pair creation \cite{BB-1991,Al-Naseri-2021,Gies2,Sheng,Kohlfurst-2020,titopar} come into play.

Before proceeding to the the more theoretical aspects, we should get an idea what concrete systems that are of interest. One such answer is explored in some detail in the recent review by Manfredi {\it et al.} \cite{Manfredi-2019}, where metallic nano-objects are in focus. As described in that work, the properties of metallic particles ranging in size from a few to several hundred nanometer can be well modeled with quantum kinetic theory. Importantly, the nano-particles are of interest for applications in nano-photonics and other technological applications. 

Furthermore, by identifying the quantum regime in a density temperature diagram, as done, e.g., by Refs. \cite{Bonitz-book, Asenjo-2012} a reasonably complete picture of quantum plasmas can be presented. Besides the solid state regime, that might be of most technological interest (including applications to e.g.,
spintronics \cite{Spintronics}, plasmonics \cite{Plasmonics}, nanotubes \cite{Nanotubes}, quantum wells \cite{Quantum-wells} and quantum dots \cite{Quantum-dots}), quantum plasmas can be found in rather diverse contexts, including dense astrophysical objects \cite{astro1,astro2} (white dwarf stars neutron stars), warm dense matter, laboratory plasmas \cite{Glenzer}, and plasmas in the early universe \cite{early universe}. However, assuming that the plasma need to occupy a specific spot in a temperature-density diagram, for quantum effects to be significant, sometimes lead to the wrong conclusion. For example, strong laser fields may potentially induce quantum phenomena such as spin polarization, as explored in Refs. \cite{SP-1,SP-2,SP-3}. Moreover, in an astrophysical context, strong magnetic fields from pulsars and/or magnetars \cite{magnetar} may lead to very pronounced Landau quantization. Finally, as we will discuss in section VI, a sufficiently strong electrostatic field lead to Schwinger pair production of electrons and positrons \cite{BB-1991,Al-Naseri-2021,Gies2,Sheng,Kohlfurst-2020}. Thus, when intense electromagnetic fields are involved, there are several ways quantum phenomena can enter the picture, even when the parameters of density and temperature correspond to the classical regime.

The paper is organized as follows: In section \ref{section2} we study foundations of  quantum kinetic theory, in particular the density matrix, the von Neumann equation, and the Wigner transform \cite{Wigner32}. To focus on the fundamentals, for much of the section we avoid the complications of spin and electromagnetic fields, and base the treatment on the simplest form of the Schrödinger equation. The theory is illustrated with examples from linear theory. 
In section \ref{section3}, the treatment is extended to cover many particle physics (in the quantum mechanical sense). In particular, the quantum mechanical version of the BBGKY-hierarchy is used to study exchange interaction, with a particular focus on electrostatic linearized theory. Exchange corrected dispersion relations are computed, for Langmuir waves and ion-acoustic waves, in the degenerate and non-degenerate regime.
In section \ref{section4}, we allow for electromagnetic fields and spin dynamics. For that purpose, the Schrödinger Hamiltonian is replaced with the Pauli Hamiltonian. Two equivalent but different versions of quantum kinetic theory is presented. The linear theory of the model is solved in the long-scale limit for the case of a homogeneous magnetized plasma, generalizing previous results to include an anisotropic background distribution.   
Next, in section \ref{section5}, a multitude of different problems are discussed that is not covered by the previous theories. This includes quantum kinetic conversation laws, relativistic Landau quantization, and aspects of nonlinear spin dynamics, focusing on the ponderomotive force. 
In section \ref{section6}, we study the fully quantum relativistic case, using the Dirac-Heisenberg-Wigner (DHW) formalism, originally derived by Ref. \cite{BB-1991}. The equations are used to study electrostatic fields, and some deviation from simpler theories are pointed out. Moreover, we show how to relate the DHW-formalism to limiting cases studied in the previous sections.
Finally, the paper ends with concluding remarks in section \ref{section7}. 

\section{The Gauge-Invariant Wigner Theory}
\label{section2}

The aim of this section is to give a brief introduction to how the Wigner functions can be used to describe quantum effects in plasmas.
In order to this, we start with a review of the density matrix and some of its properties. 
After that, we go on to define the Wigner transformation, which is a transformation that brings the evolution equation in a form similar to classical kinetic theory. 
We comment on the interpretation of the Wigner function and some of the key properties. 
Finally, we consider a system in an electromagnetic-field and give a description of the gauge-invariant Wigner function and the corresponding evolution equation. 

\subsection{The Density Matrix} 
In order to describe an $N$-particle system, where the initial state is known we can use a many-particle wave function $\psi = \psi(\mathbf{r}_{1}, \dots \mathbf{r}_N, t)$.
Here $\mathbf{r}_i$ is the position for the $i$:th particle and $\left| \psi(\mathbf{r}_1, \dots \mathbf{r}_N), t \right|^2 d^3r_1 \dots d^3r_N $,
is the probability of simultaneously measuring the particles at respective positions in the volume $d^3r_i$ around $\mathbf{r}_i$. 
Assuming that the particles have mass $m$ and are moving in a potential $V = V(\mathbf{r}_1, \dots \mathbf{r}_{N})$, the evolution follows the Schrödinger equation
\begin{equation}
	i \hbar \partial_t \psi ( \mathbf{r}_1, \dots \mathbf{r}_N , t)  
    = \hat{H} \psi 
    = - \frac{\hbar^2}{2m} \sum_{i = i}^N \nabla_{\mathbf{r}_i}^2 \psi + V \psi , 
\end{equation} 
supplemented by the initial state of the system 
\begin{equation}
    \psi( \mathbf{r}_1, \dots \mathbf{r}_N , t = 0) = \psi_0 (\mathbf{r}_1, \dots \mathbf{r}_N ) . 
\end{equation}
The exact state of a multi-particle system cannot be known in general, instead the best we can hope for is the knowledge of a statistical distribution of states.
In order to handle this situation in it is then necessary to use the so called density matrix. 
It is defined as 
\begin{equation}
    \rho(\mathbf{r}_1, \dots \mathbf{r}_N , \mathbf{r}'_1, \dots \mathbf{r}'_N ,t ) 
    = 
    \sum_i p_i \psi_i (\mathbf{r}_1, \dots \mathbf{r}_N, t) \psi^*(\mathbf{r}'_1, \dots \mathbf{r}'_N, t), 
\end{equation}
where $p_i$ is the probability of finding the system in state $\psi_i$, and the sum is over all states that the system can be in. 
From normalisation we require that 
\begin{equation}
    \sum_i p_i = 1 ,
\end{equation}
which can be interpreted that the total probability of finding the system in one of the states $\psi_i$ is unity. 
E.g., in the case where the system is in thermal equilibrium with temperature $T$, the density matrix is
\begin{equation}
    \rho( \mathbf{r}_1, \dots \mathbf{r}_N , \mathbf{r}'_1 , \dots \mathbf{r}'_N ) 
    = \frac{1}{Z} \sum_i e^{-E_i / k_B T} 
    \psi_{E_i} (\mathbf{r}_1, \dots \mathbf{r}_N) 
    \psi_{E_i}^*(\mathbf{r}'_1, \dots \mathbf{r}'_N) ,  
\end{equation}
where  
\begin{equation}
    Z = \sum_i e^{-E_i / k_B T} 
\end{equation}
is the partition function, and the sum is over all eigenstates $\psi_{E_i}$ to the Hamiltonian, $\hat{H} \psi_{E_i} = E_i \psi_{E_i}$.

From the definition above together with the Schrödinger equation (and its complex conjugate), it is we can derive the evolution equation for the density matrix
\begin{multline}
    i \hbar \partial_t \rho(\mathbf{r}_1, \dots \mathbf{r}_N, \mathbf{r}'_1, \dots \mathbf{r}'_N, t) 
    =
    - \frac{\hbar^2}{2m} \sum_i \left( \nabla_{\mathbf{r}_i}^2 - \nabla_{\mathbf{r}'_i}^2 \right)
    \rho(\mathbf{r}_1, \dots \mathbf{r}_N, \mathbf{r}'_1, \dots \mathbf{r}'_N, t) 
    \\
    + \left[ V( \mathbf{r}_1, \dots \mathbf{r}_N) - V(\mathbf{r}'_1, \dots \mathbf{r}'_N) \right] 
    \rho(\mathbf{r}_1, \dots \mathbf{r}_N, \mathbf{r}'_1, \dots \mathbf{r}'_N, t) .
	\label{vonneumannpos}
\end{multline}
This is called the von Neumann equation, or alternatively, the quantum Liouville equation (see e.g., Ref.~\cite{Bonitz-book}). 
The density matrix describes the system completely, and we can use it to calculate, e.g., the probability of finding the particles at respective positions $\mathbf{r}_1, \dots \mathbf{r}_N$ as the diagonal elements
\begin{equation}
    \rho( \mathbf{r}_1, \dots \mathbf{r}_N , \mathbf{r}_1, \dots \mathbf{r}_N , t ) 
    = \sum_i p_i \left| \psi(\mathbf{r}_1, \dots \mathbf{r}_N, t) \right|^2 . 
\end{equation}
One thing to note about the density matrix is that it changes under a gauge-transformation. 
E.g., in case of a single particle interacting with an electromagnetic field described by the potentials $\phi$, $\mathbf{A}$, a gauge transformation 
\begin{align}
    \phi(\mathbf{r},t) & \rightarrow \phi(\mathbf{r},t) - \partial_t \Lambda(\mathbf{r},t) \\
    \mathbf{A} (\mathbf{r},t) & \rightarrow \mathbf{A} (\mathbf{r},t) + \nabla \Lambda (\mathbf{r},t)
\end{align}
the density matrix transforms as 
\begin{equation}
    \rho(\mathbf{r}, \mathbf{r}', t) \rightarrow e^{iq \Lambda(\mathbf{r}, t) / \hbar} 
    \rho(\mathbf{r} , \mathbf{r}' , t) e^{-i q\Lambda (\mathbf{r}', t) / \hbar },
	\label{gauge-density-matrix}
\end{equation}
where $q$ is the charge of the particle, see e.g., \cite{Sakurai-book}. 
In the $N$-particle case, the density matrix obtains a phase-factor $\exp\left(i q \Lambda (\mathbf{r}_i ,t ) /\hbar \right)$ for each $\mathbf{r}_i$, and a factor $\exp\left(-iq \Lambda(\mathbf{r}'_i ,t) / \hbar \right)$ for each $\mathbf{r}_i'$.

The density matrix encodes all the information about the system. 
However, the evolution equation \eqref{vonneumannpos} involves a huge number of variables, six for each particle, so it is not directly applicable when considering plasmas. 
In Appendix~\ref{appendix}, we derive the mean-field approximation, which is a common approximation used when modelling plasmas. 
We here state the most important conclusions. 
In the mean-field approximation density matrix only depends on two variables 
 \begin{equation}
	 \rho (\mathbf{r}, \mathbf{r}', t) ,
\end{equation}
where the diagonal elements yields the density of particles at a given position in space
\begin{equation}
	n(\mathbf{r}, t) = \rho(\mathbf{r}, \mathbf{r}, t), 
\end{equation}
with the normalisation
\begin{equation}
	\int d^3 r \rho(\mathbf{r}, \mathbf{r}, t) = N ,
\end{equation}
where $N$ is the total number of particles. 
The mean-field Hamiltonian formally looks like it is describing a one-particle system, but is modified in that the fields are the self-consistent fields created by all the particles. 
For example, for an electrostatic plasma we have 
\begin{equation}
	\hat H = \frac{\hat{\mathbf{p}}^2}{2m} + q \phi (\mathbf{r}, t) ,
	\label{mfham} 
\end{equation}
where $\phi$ is the self-consistent field, which is relate to the particle density via Poisson's equation
\begin{equation}
	\nabla^2 \phi (\mathbf{r},t ) = - \frac{q}{\epsilon_0} \rho(\mathbf{r}, \mathbf{r}, t) + \frac{q n_0}{\epsilon_0} .
\end{equation}
Here we have added a neutralising and homogeneous charge density $- q n_0$. 
The mean-field approximation is usually applicable in cases where particle correlations can be neglected. 
E.g., in cases where particle-particle collisions can be neglected. 
By including further terms in the BBGKY-hierarchy it is possible to include particle collisions, see e.g., Ref.~\cite{Bonitz-book}  
In the discussion above we have also neglected any reference to the particle statistics.
Since we are dealing with fermions one should really take into account the anti-symmetry of the wave function. 
In Section \ref{section3} we generalise the mean-field approximation to account for this. 
The resulting approximation is usually called the Hartree-Fock approximation. 

\subsection{The Wigner Transformation} 
\label{sec2b}
A key tool when deriving quantum kinetic equations from the Schrödinger or Pauli Hamiltonian is the Wigner transform, introduced by Eugene Wigner in his famous paper from 1932 \cite{Wigner32}. Using this transform on the density matrix, a kinetic evolution generalizing the classical Vlasov equation can be derived. This connection with classical theory is helpful for guiding the physical intuition, and the transformed quantities are typically more attractive for practical calculations.  Wigner's original approach has been further developed by others, in particular José Enrique Moyal \cite{{Moyal-1949}}, and the Wigner equation is sometimes also referred to as the Wigner-Moyal equation. 

The Wigner transform of the density matrix is defined by 
\begin{equation}
	W(\mathbf{r}, \mathbf{p}, t) 
	= 
	\int \frac{d^3 z}{\left( 2 \pi \hbar \right)^3}
	e^{- i \mathbf{z} \cdot \mathbf{p} / \hbar} 
	\rho( \mathbf{r} + \mathbf{z} / 2 , \mathbf{r} - \mathbf{z} / 2, t) .
	\label{wigner-def}
\end{equation}
We can see that e.g., 
\begin{equation}
	\int d^3 p W(\mathbf{r}, \mathbf{p}, t) = \rho ( \mathbf{r}, \mathbf{r} , t) = n(\mathbf{r}, t) 
\end{equation}
gives the local density of particles, and similarly
\begin{equation}
	\int d^3 r W(\mathbf{r}, \mathbf{p}, t) = n(\mathbf{p}, t)
\end{equation}
is the momentum distribution of the particles. 
However, in contrast to the Vlasov distribution, the Wigner distribution can be negative in certain regions. 
This is attributed to the Pauli exclusion principle, and we should hence be careful in interpreting the Wigner function $W(\mathbf{r}, \mathbf{p}, t)$ as a classical phase-space distribution. 
Irrespective of that, the Wigner function can still be used to calculate macroscopic properties of the system, such as, e.g., the charge and current densities. 

If the system is described by the Hamiltonian, \eqref{mfham}, we can derive the evolution equation for $W$ using the Schrödinger equation as follows: 
The evolution equation for the density matrix is given by 
\begin{equation}
	i \hbar \partial_t \rho(\mathbf{x} , \mathbf{x}' , t) = 
	-\frac{\hbar^2}{2m} \left( \nabla_\mathbf{x}^2 - \nabla_{\mathbf{x}'}^2 \right) \rho(\mathbf{x}, \mathbf{x}' , t) 
	+ q \left[ \phi(\mathbf{x}, t) - \phi(\mathbf{x}', t) \right] \rho(\mathbf{x}, \mathbf{x}', t) .
\end{equation}
Using the variable change
\begin{equation}
	\begin{cases}
		\mathbf{r} = \frac{\mathbf{x} + \mathbf{x}'}{2} \\
		\mathbf{z} = \mathbf{r} - \mathbf{r}' 
	\end{cases}
\end{equation}
the equation is cast into 
\begin{multline}
	i \hbar \partial_t \rho( \mathbf{r} + \mathbf{z} / 2, \mathbf{R} - \mathbf{z} /2 ,t ) 
	=
	- \frac{\hbar^2}{m} \nabla_\mathbf{r} \cdot \nabla_\mathbf{z}
	\rho( \mathbf{r} + \mathbf{z} / 2, \mathbf{r} - \mathbf{z} /2 ,t ) 
	\\ 
	+ q \left[ \phi( \mathbf{r} + \mathbf{z} / 2) - \phi ( \mathbf{r} - \mathbf{z} / 2) \right]
	\rho( \mathbf{r} + \mathbf{z} / 2, \mathbf{r} - \mathbf{z} /2 ,t ) .
\end{multline}
We now want to take the Fourier transformation of this and re-identify the Wigner function in each term. 
In order to do this, we can use the following trick 
\begin{multline}
	\int d^3z e^{- i \mathbf{p} \cdot \mathbf{z} / \hbar} f(\mathbf{r}, \mathbf{z}) 
	\rho(\mathbf{r} + \mathbf{z} / 2, \mathbf{r} - \mathbf{z} / 2) 
	=
	\int d^3z e^{- i \mathbf{p} \cdot \mathbf{z} / \hbar} f(\mathbf{r}, i \hbar \overleftarrow{\nabla}_\mathbf{p}) 
	\rho(\mathbf{r} + \mathbf{z} / 2, \mathbf{r} - \mathbf{z} / 2)
	\\
	= f(\mathbf{r} ,  i \hbar \overrightarrow{\nabla}_\mathbf{p}) \int d^3z e^{- i \mathbf{p} \cdot \mathbf{z} / \hbar} 
	\rho(\mathbf{r} + \mathbf{z} / 2, \mathbf{r} - \mathbf{z} / 2) , 
\end{multline}
where $f(\mathbf{r}, i \hbar \nabla_\mathbf{p})$ is defined by its Taylor expansion. 
Here the arrow over $\overleftarrow \nabla_\mathbf{p}$ indicates that the operator should act to the left on the exponential function only. 
For the term containing $\nabla_\mathbf{z}$ we make a partial integration. 
Doing this we finally get the equation
\begin{equation}
	\partial_t W(\mathbf{r}, \mathbf{p}, t) 
	= 
	- \frac{1}{m} \mathbf{p} \cdot \nabla_\mathbf{r} W( \mathbf{r}, \mathbf{p}, t) 
	- \frac{i q}{\hbar} 
	\left[ \phi( \mathbf{r} + i \hbar \nabla_\mathbf{p} / 2) - \phi( \mathbf{r} - i \hbar \nabla_\mathbf{p} / 2 ) \right] 
	W( \mathbf{r}, \mathbf{p}, t ) .
	\label{wigner-vlasov}
\end{equation}
In the mean-field approximation, the equation above describes a system of particles interacting via the electric potential created by all the particles via  
\begin{equation}
	\nabla^2 \phi(\mathbf{r}, t)
	= - \frac{q}{\epsilon_0} \left( n(\mathbf{r}, t) - n_0 \right)
	= - \frac{q}{\epsilon_0} \int d^3 p W(\mathbf{r}, \mathbf{p}, t) + \frac{qn_0}{\epsilon_0} ,
\end{equation}
where we have assumed that there is a homogeneous, neutralizing background of particles with charge density $-qn_0$ (note, $q = -e < 0$ for electrons with our conventions). 
These two equations together with the appropriate initial conditions yields the dynamics of the system. 

The Wigner distribution $W$ function, can be used to calculate any expectation value in a similar way to how this is done in classical kinetic theory. For example we have 
\begin{equation}
	\left< \mathbf{r} \right> = \int d^3 r d^3 p \mathbf{r} W(\mathbf{r}, \mathbf{p}, t). 
\end{equation}
For a general operator $\hat{O} = O(\hat{\mathbf{r}}, \hat{\mathbf{p}})$, depending on both the position and momentum variables, we must first put all the position and momentum operators in completely symmetric form using the commutation relations, and then replace all operators with the corresponding phase-space variables. 
For example, to calculate the expectation value of the operator $\hat{O} = \hat{x} \hat{p}_x \hat{x}$ we first write 
\begin{equation}
	\hat{O} = \frac{1}{3} \left( \hat{x} \hat{p}_x \hat{x} + \hat{x} \hat{p}_x \hat{x} + \hat{x} \hat{p}_x \hat{x} \right) = 
	\frac{1}{2} \left( \hat{x} \hat{p}_x + \hat{p}_x \hat{x} + i \hbar \right) 
	\rightarrow x p_x + i \hbar. 
\end{equation}
So the phase-space function corresponding to $\hat{O}$ is $O = x p_x + i \hbar$. 
Note, that this particular operator is not physical since it is not hermitian, we only chose it to illustrate the procedure. 
This correspondence between operators and phase-space functions is called \textit{Weyl ordering}, see e.g., \cite{qmps}. 

As an example, consider linear waves where the distribution function can be written
\begin{equation}
	W (\mathbf{r}, \mathbf{p}, t) = W_0 (\mathbf{p}) + W_1(\mathbf{p}) e^{ikz - i\omega t}  
\end{equation}
where $W_0 (\mathbf{p})$ is a spatially homogeneous equilibrium, e.g., a Maxwell-Boltzmann distribution. 
Similarly we write the potential as 
\begin{equation}
	\phi(\mathbf{r}, t) = \phi_1 e^{ikz - i\omega t} .  
\end{equation}
Linearising the evolution equations above and solving for $\phi_1$ we get the dispersion relation 
\begin{equation}
	1 + \frac{q^2}{\epsilon_0 \hbar k^2} \int d^3 p 
	\frac{W_0 (\mathbf{p} + \hbar k \hat{\mathbf{z}}/2) - W_0 ( \mathbf{p} - \hbar k \hat{\mathbf{z}} / 2) }{\omega - k p_z / m }  
	=
	1 - \frac{q^2}{m\epsilon_0}
	\int d^3 p \frac{W_0 (\mathbf{p}) }{\left(\omega - k p_z / m \right)^2 - \hbar^2 k^4 / 4 m^2}  
	= 0 , 
\end{equation}
where the last equation is obtained by a change of variables. 
In the limit $\hbar \to 0$, this reduces to the classical Vlasov dispersion relation. 
This dispersion relation has been investigated in detail in Ref. \cite{eliason-shukla-09}. 
There it is shown that for a fully degenerate background distribution ($T = 0$), the wave particle damping disappears in the Vlasov limit, since the phase-velocity $\omega / k$ always exceeds the Fermi velocity. 
Furthermore, the critical wave number $k_c$, at which the Landau damping sets in, was computed. 

Returning to the general case, Eq.~\eqref{wigner-vlasov}, it is straightforward to show that this equation reduces to the classical Vlasov equation for long macroscopic scale lengths. In particular, if the potential varies on a scale length $L$, i.e. $\nabla \phi \sim \phi / L$, and the characteristic velocity of the system is  $v$, $\nabla_p W \sim W / m v$, then in the limit
\begin{equation}
	L \gg \frac{\hbar}{m v} ,
	\label{semiclasslim}
\end{equation}
we may keep the first non-vanishing terms in a Taylor expansion of the potential in Eq.~\eqref{wigner-vlasov}. 
We then get 
\begin{equation}
	\partial_t W + \frac{\mathbf{p}}{m} \cdot \nabla_\mathbf{r} W + q \mathbf{E} \cdot \nabla_\mathbf{r} W = 0, 
\end{equation}
i.e., the Vlasov equation, where $\mathbf{E} = - \nabla_\mathbf{r} \phi$.
By keeping further terms in the expansion we may use this method to derive quantum corrections to an arbitrary order, see, e.g., Ref.~\cite{Waterbag}. 

\subsection{The Gauge-Invariant Wigner Function}\label{IIC}
The Hamiltonian describing a charged particle of mass $m$ and charge $q$ ($q = -e$ for electrons) interacting with a magnetic field is given by 
\begin{equation}
    \hat H = \frac{\left[ \hat{\mathbf{p}} - q \mathbf{A}(\mathbf{r}, t) \right]^2}{2m} + q \phi(\mathbf{r}, t) ,  
\end{equation}
where $\phi$ and $\mathbf{A}$ are the (mean-field) electromagnetic potentials. 
Note that, here $\hat{\mathbf{p}} = - i \hbar \nabla_\mathbf{r}$ is the canonical momentum operator, and it is related to the velocity via the vector potential in the usual way. 
If the state of the particle is described by the density operator $\hat \rho (t)$, the evolution of the system is given by the von Neumann equation~\eqref{vonneumann1}. As we have seen, above, the equation can also be used to describe a system of particles interacting with the mean-field created by all the other particles. 
In that case, the density operator should be interpreted as the reduced density operator. 
This is the point of view we will take here. 

Under a gauge transformation the density matrix changes according to Eq.~\eqref{gauge-density-matrix}. 
Due to this, the Wigner function $W$, defined in Eq.~\eqref{wigner-def} will not be gauge-invariant. 
Furthermore, the momentum variable, $\mathbf{p}$ will be related to the velocity in a gauge-dependent way. 
In order to get something that is more attractive to work with, we will use a modified version of the Wigner transformation, first constructed by Stratonovich \cite{strato}. 
The definition is 
\begin{equation}
	W(\mathbf{r}, \mathbf{p}, t) 
	= 
	\int 
	\frac{d^3 z}{( 2 \pi \hbar )^3}
	\exp\left( - \frac{i}{\hbar} \mathbf{p} \cdot \mathbf{z} 
	- \frac{iq}{\hbar} \mathbf{z} \cdot \int_{-1 / 2}^{1 / 2} d\lambda \mathbf{A} (\mathbf{r} + \lambda \mathbf{z}, t ) \right) 
	\rho( \mathbf{r} + \mathbf{z} / 2 , \mathbf{r} - \mathbf{z} / 2 , t) .
	\label{wigstrat}
\end{equation}
An important aspect of this transformation, is that the momentum variable $\mathbf{p}$, is the \textit{kinetic momentum} related to the velocity via $\mathbf{p} = m \mathbf{v}$.
This can be seen by calculating the momentum density 
\begin{equation}
	\mathbf{P} ( \mathbf{r} , t) 
	\equiv 
	\int d^3 p \mathbf{p} W(\mathbf{r}, \mathbf{p}, t)  
	= 
	\left.
	\frac{1}{2} 
	\left[ 
		- i\hbar \nabla_\mathbf{r} - q \mathbf{A}(\mathbf{r}, t) + i \hbar \nabla_\mathbf{r'} - q \mathbf{A} (\mathbf{r}' ,t) 
	\right] 
	\rho( \mathbf{r} , \mathbf{r}', t) \right|_{\mathbf{r}' = \mathbf{r}}
	= \frac{1}{2} \left< \mathbf{r} \right| \left\{ \hat{\mathbf{p}} - q \mathbf{A} (\hat{\mathbf{r}} , t) , \rho 
	\right\} 
	\left| \mathbf{r} \right> , 
\end{equation}
where $\left\{ \cdot , \cdot \right\}$ denotes the anti-commutator. 
The right hand side can be identified as the (kinetic) momentum density. 
From the evolution equation for the density matrix we can derive the evolution equation in the gauge invariant case in a similar fashion as in the previous sub-section. The result is~\cite{serimaa} 
\begin{equation}
	\partial_t W + \frac{1}{m} \left( \mathbf{p} + \Delta \mathbf{p} \right) \cdot \nabla_\mathbf{r} W 
	+ q \left[ \tilde{\mathbf{E}} + \frac{\mathbf{p}}{m} \times \tilde{\mathbf{B}} \right] \cdot \nabla_\mathbf{p} W 
	= 0 ,
	\label{gaugeevolu}
\end{equation}
where 
\begin{align*}
	\Delta \mathbf{p} (\mathbf{r}, t) &= - iq \hbar 
	\int_{-1 / 2}^{1 / 2} d\lambda 
	\mathbf{B} \left( \mathbf{r} + i \hbar \nabla_\mathbf{p}, t \right) \cdot \nabla_\mathbf{p}
	\\
	\tilde{\mathbf{E}} (\mathbf{r} , t ) &= 
	\int_{-1 / 2}^{1 / 2} d\lambda 
	\mathbf{E} \left( \mathbf{r} + i \hbar \nabla_\mathbf{p}, t \right) \cdot \nabla_\mathbf{p}
	\\
	\tilde{\mathbf{B}} (\mathbf{r} , t ) &= 
	\int_{-1 / 2}^{1 / 2} d\lambda 
	\mathbf{B} \left( \mathbf{r} + i \hbar \nabla_\mathbf{p}, t \right) \cdot \nabla_\mathbf{p} . 
\end{align*}
We note that, since the equation is completely expressed in terms of the electric and magnetic field, it is manifestly gauge-invariant. 
Also, since the momentum variable involved is the kinetic momentum, we may make the trivial variable change $\mathbf{p} = m \mathbf{v}$ and express the Wigner function as $W = W (\mathbf{r}, \mathbf{v}, t)$.
In the classical limit, i.e. when Eq. \eqref{semiclasslim} applies, this reduces to the Vlasov equation. 
By Taylor-expanding the functions $\Delta \mathbf{p}, \tilde{\mathbf{E}}$, and $\tilde{\mathbf{B}}$, it is possible to obtain approximations to arbitrary order in $\hbar$. 

The models we have explored in this section are of importance in their own right, and have been used, for example, to consider quantum dispersive effects, see e.g., Refs.~\cite{Shukla-Eliasson-2011, Vladimirov-review} and references therein. 
However, in what follows, we will continue with more elaborate models including additional physical phenomena, such as exchange effects, spin, and relativistic effects.

\section{Exchange effects}
\label{section3}

Exchange interaction in plasmas follows from electrons being fermions, with a totally anti-symmetric wavefunction. To include exchange effects it is not enough to just use the Slater determinants to construct the many-body wave-functions from single-particle wave-functions (although this is a necessary first step). Nor will it suffice to simply infer Fermi-Dirac statistics of the background plasma. Instead one must go go beyond the simplest mean-field description. Covering exchange effects is usually referred to as the Hartee-Fock approximation \cite{Hartree-Fock}, in contrast to the Hartree approximation, where exchange effects are ignored.

The relative importance of exchange effects in plasmas is proportional to the parameter $H^2=(\hbar\omega_p/E_k)^2$ \cite{Manfredi-exchange,Haas-exchange,Egen-2013-exchange,Egen-2015-exchange,Egen-2015-second-exchange,Egen-2019-exchange}, where the characteristic kinetic energy $E_k$ is given by $E_k=k_B T_F$ for a degenerate plasma (with $T_F>T$) and by $E_k=k_B T$ for the non-degenerate case (with $T>T_F$). The above suggests that exchange effects are as important as the more basic particle dispersive effects, described by the Wigner-Moyal equation of the previous section. If so, the common approach of including particle dispersive effects through the Wigner equation, but simultaneously neglecting exchange effects would be highly questionable. Fortunately, while the relative importance scales with temperature and density as given by the $H$-parameter, the overall importance is also proportional to another dimensionless constant that often is much smaller than unity \cite{Egen-2019-exchange}. Hence the use of the Wigner equation from the preceding section, neglecting exchange effects, can still be a good approximation. What complicates the picture, is that the importance of exchange effects it not just dependent on the background plasma parameters (temperature and density), but also of the specific problem under study. Below we will illustrate this by considering simple examples of high-frequency Langmuir waves, and low-frequency ion-acoustic waves.

\subsection{General electrostatic theory}

Since the theory becomes more complicated for the electromagnetic case, we here present a general quantum kinetic approach to exchange effects in the electrostatic limit. The full electromagnetic case will be discussed briefly in the end of Section III.  

Our treatment will follow Ref. \cite{Egen-2013-exchange}, but leaving out some of the technical details. We here consider a completely ionized electron-ion plasma with
the particles interacting through a mean-field scalar potential. Quantum
effects for the ions will be completely neglected  as will effects due to the self-energy and
particle correlations \cite{Bonitz-book}. 
The state of the $N$-electrons is described by the density operator 
$\hat{\rho}_{1\dots N}$ (see for example Ref.\ \cite{Bonitz-book}), and the dynamics is
given by the von Neumann equation with the Hamiltonian 
\begin{equation}
\hat{H}_{1\dots N}=\sum_{i=1}^{N}\frac{\hat{\mathbf{p}}_{i}^{2}}{2m_{e}}+\frac{e^{2}}{%
4\pi \epsilon _{0}}\sum_{i<j}\frac{1}{|\hat{\mathbf{x}}_{i}-\hat{\mathbf{x}}%
_{j}|}+e\sum_{i=1}^{N}\varphi (\hat{\mathbf{x}}_{i}).
\end{equation}%
The last term accounts for the
interaction with the electric potential created by the ions. 
In Appendix \ref{appendix} we derive the mean-field approximation, in the case where we can neglect exchange effects.
Here we generalize \eqref{reduced1} to take into account the anti-symmetry of the wave function for the electrons.  
We introduce the reduced density operators according to 
\begin{equation}
\hat{\rho}_{1\dots i} = N^s \mathrm{Tr}_{s+1\dots N}\hat{\rho}_{1\dots N}\hat{%
\Lambda}_{1\dots i},
\end{equation}%
where $\mathrm{Tr}_{s+1\dots N}$ denotes the trace over particles $s+1$ to $N$ 
(i.e. integrating over the position degree of freedom and summing over
the spins), and $\hat{\Lambda}_{1\dots s}$ is the
anti-symmetrization operator that takes an $s$-particle state and makes it
completely anti-symmetric \cite{db79}. We will only need to know that $\hat{%
\Lambda}_{12}=1-\hat{P}_{12}$ where $\hat{P}_{12}$ interchanges particle 1
and 2, i.e. $\hat{P}_{12}\psi (\mathbf{x}_{1},\mathbf{x}_{2})=\psi (\mathbf{x%
}_{2},\mathbf{x}_{1})$ (see, e.g., Ref.\ \cite{Bonitz-book} for further details).
The evolution for the one-particle density operator is given by 
\begin{equation}
i\hbar \partial _{t}\hat{\rho}_{1}=[\hat{h}_{1},\hat{\rho}_{1}] + \mathrm{Tr}%
_{2}[\hat{V}_{12},\hat{\rho}_{12}\hat{\Lambda}_{12}],  \label{rho1}
\end{equation}%
where $\hat{h}_{1}=\hat{\mathbf{p}}^{2}/(2m_{e})$ and $\hat{V}_{12}=V(\hat{\mathbf{x}}%
_{1}-\hat{\mathbf{x}}_{2})=e^{2}/(4\pi \epsilon _{0}|\hat{\mathbf{x}}_{1}-%
\hat{\mathbf{x}}_{2}|)$ and $\hat{\rho}_{12}$ is the two-particle density
operator. The effects of two-particle correlations $\hat{g}_{12}$ can be
separated out of the two-particle density operator by writing it in the form 
\begin{equation}
\hat{\rho}_{12}=\hat{\rho}_{1}\hat{\rho}_{2}+\hat{g}_{12},
\label{correlations}
\end{equation}%
see e.g., Ref.\ \cite{wang85}. We are interested in the collisionless limit
where a mean-field approximation will suffice. This approximation is
obtained by neglecting the correlation $\hat{g}_{12}$. Utilizing this in Eq.~\eqref{rho1}
we obtain 
\begin{equation}
i\hbar \partial _{t}\hat{\rho}_{1}=[\hat{h}_{1},\hat{\rho}_{1}]+[\bar{V}_{1},%
\hat{\rho}_{1}],  \label{rhomf}
\end{equation}%
where $\bar{V}_{1}=\mathrm{Tr}_{2}\hat{V}_{12}\hat{\rho}_{2}\hat{\Lambda}%
_{12}$, is the Hartree-Fock potential operator. This is a closed system for
the one-particle density operator.

To obtain a connection to the classical kinetic theory we use the Wigner
representation \cite{Wigner32} of this equation. Using the complete set of
states $\left| \mathbf{x}, \alpha \right>$, where $\mathbf{x}$ is the
position and $\alpha = 1,2$ is the spin along the axis of quantization, this
representation is obtained as
\begin{eqnarray}
f(\mathbf{x}, \mathbf{p}, \alpha, \beta ) &=& \frac{1}{(2\pi\hbar)^{3} }
\int d^3\! y \,\, e^{i \mathbf{y }\cdot \mathbf{p }/ \hbar} \rho \left( 
\mathbf{x }+ \frac{\mathbf{y}}{2} , \alpha ; \mathbf{x }- \frac{\mathbf{y}}{2%
} , \beta \right), 
\end{eqnarray}
where $\rho(\mathbf{x}, \alpha ; \mathbf{y }, \beta) = \left< \mathbf{x},
\alpha \right| \hat \rho_1 \left| \mathbf{y}, \beta \right>$ is the density
matrix. 
This is a slight generalisation of Eq.~\eqref{wigner-def}, where we trivially include the spin-variables. 
Note, however, that the resulting Wigner function depends on the two spin-variables and is hence a 
2-by-2 matrix. 
Writing Eq.\ \eqref{rhomf} first in the position representation and
Wigner transforming the result (using e.g., the method outlined in Section \ref{sec2b}) we obtain, 
\begin{eqnarray}
	\partial_t  f ( \mathbf x , \mathbf p , \alpha, \beta) 
	+ \frac{1}{m} \mathbf p \cdot \nabla_\mathbf{x} f(\mathbf x , \mathbf p, \alpha, \beta)    
	+ \frac{i e}{\hbar} \int  \frac{ d^3\! y \, d^3\! p'}{(2 \pi \hbar)^3} \, 
	e^{i \mathbf y \cdot ( \mathbf p - \mathbf p')/ \hbar} 
	\left[ \phi \left( \mathbf x + \frac{\mathbf y}{2} \right) 
		- \phi \left( \mathbf x - \frac{\mathbf y}{2} \right) \right] 
	f( \mathbf x, \mathbf p' , \alpha,  \beta) 
		\notag \label{14} \\
	=
	  \frac{i}{\hbar(2\pi \hbar)^3} \sum_{\gamma=1}^2 \int d^3\! p' \, d^3\! p'' \, d^3\! y \, d^3\! r \,\,
	 e^{ i \mathbf p \cdot \mathbf y/ \hbar } 
	 e^{ - i \mathbf p' \cdot ( \mathbf x + \mathbf y /2 - \mathbf r)/\hbar} 
	 e^{ - i \mathbf p'' \cdot ( \mathbf r - \mathbf x + \mathbf y /2 ) / \hbar} 
	 \notag \\
	 \times \left[ V \left( \mathbf x + \frac{\mathbf y}{2} - \mathbf r \right) 
	 - V \left( \mathbf x - \frac{\mathbf y}{2} - \mathbf r \right) \right] 
	 f \left( \frac{\mathbf x + \mathbf r}{2} + \frac{\mathbf y}{4} , \mathbf p' , \alpha, \gamma \right) 
	 f \left( \frac{\mathbf x + \mathbf r}{2} - \frac{\mathbf y}{4} , \mathbf p'' , \gamma, \beta \right) ,
\end{eqnarray} 
where 
\begin{equation}
\phi (\mathbf{x}) = \frac{e n}{4 \pi \epsilon_0} \sum_{\gamma=1}^2 \int
d^3\! z \,\, \frac{\rho(\mathbf{z }, \gamma; \mathbf{z} ,\gamma )}{| \mathbf{%
x }- \mathbf{z }|} + \varphi(\mathbf{x}).
\end{equation}
is the total (mean-field and the ionic field) potential and 
\begin{equation}
V(\mathbf{x}) = \frac{e^2}{4\pi \epsilon_0 |\mathbf{x}| }
\end{equation}
is the Coulomb potential. The left hand side of Eq.\ \eqref{14} represents
the electrostatic limit of the Wigner-Moyal equation, but keeping the spin-dependence (as encoded in $(\alpha, \beta)$) while the right hand side is the correction due
to exchange effects. This term is nonlocal in phase-space and nonlinear in
the distribution function.

Two steps of our general treatment remain, taking the long-scale limit and averaging over the spin states $(\alpha, \beta)$. The first step is straightforward, just expanding the equations in $\hbar\nabla_x\nabla_p$. The second step takes a little more work, but can be done with the help of the spin transform. The spin transform will be discussed in detail in the next section (Section IV), and the details regarding the procedure for our specific case can be found in Ref. \cite{Egen-2013-exchange}. Thus we will omit these details here, and proceed directly to the end result, the spin averaged evolution equation (assuming all spin directions equally probable) in the long scale limit ($\hbar\nabla_x\nabla_p\ll 1$). The evolution equation then reads 
\begin{multline} 
	\partial_t f (\mathbf x, \mathbf p, t) 
	+ \frac{\mathbf p}{m} \cdot \nabla_\mathbf{x} f(\mathbf x, \mathbf p, t) 
	+ e \mathbf E (\mathbf x,t) \cdot \nabla_\mathbf{p} f(\mathbf x ,\mathbf p, t) 
	 =
    \\
	 \frac{1}{2} \partial_p^i \int d^3\! r \, d^3\! p' \,\, e^{ - i \mathbf r \cdot \mathbf p' / \hbar} 
	[\partial_r^i V (\mathbf r)] 
	f \left( \mathbf x - \frac{\mathbf r}{2} , \mathbf p + \frac{\mathbf p'}{2}, t \right) 
	f \left( \mathbf x - \frac{\mathbf r}{2} , \mathbf p - \frac{\mathbf p'}{2}, t \right)  
    \\ 
	- \frac{i \hbar}{8}  
	\partial_p^i \partial_p^j \cdot \int d^3\! r \, d^3\! p' \,\, e^{ - i \mathbf r \cdot \mathbf p' / \hbar} 
	[\partial_r^i V (\mathbf r) ]
    \\ 
	\times
	\left[ f \left( \mathbf x - \frac{\mathbf r}{2} , \mathbf p - \frac{\mathbf p'}{2},  t \right) 
	\left( \overleftarrow \partial_x^j  - \overrightarrow \partial_x^j \right)
	f \left( \mathbf x - \frac{\mathbf r}{2} , \mathbf p + \frac{\mathbf p'}{2} , t \right)  \right]
\label{23}
\end{multline} 
where $\partial_x^i \equiv \partial/\partial x_i$ and analogously for $%
\partial_p^i$ and an arrow above an operator indicates in which direction it
acts. We have also used the summation convention so that a sum over indices
occurring twice in a term is understood.
\subsection{High-frequency Langmuir waves}
Eq. (\ref{23}) is derived using a perturbative approach where exchange effects are considered to be small. Thus there is little reason trying to solve Poisson's equation together with (\ref{23}) exactly. Instead we first solve Eq. (\ref{23}) dropping the right hand side altogether (i.e. we solve the Vlasov equation), and then substitute these solutions into the right hand side, in order to evaluate the exchange correction to first order. Even this simplified treatment might be rather difficult, unless the zero order Vlasov solution is fairly simple. In order to focus on a case that can be treated analytically to a large degree, we now consider the case of linear Langmuir waves in a homogeneous plasma. The general procedure is as follows: 
\begin{enumerate}
\item Pick a background distribution function (e.g. a Maxell-Boltzmann or a Fermi-Dirac distribution), linearize the left hand side Vlasov equation, make a plane wave ansatz, and compute the perturbed distribution function. 
\item Linearize the right hand side exchange term (keep terms where one factor is the linear perturbation, and the other is the background), substitute the distribution functions from the previous step and compute the integrals, and find the exchange correction to the perturbed distribution function.  
\item Make a final momentum integration to find the exchange contribution to the charge density, and use this in Poisson's equation to find the exchange correction to the susceptibility. 
\end{enumerate}
Treating the ions as immobile, considering a Maxwell-Boltzmann background distribution for the electrons, and performing the steps outlined above, Ref. \cite{Egen-2015-exchange} was able to derive the following result: 

\begin{equation}
1=\chi _{\mathrm{cl}}+\chi _{\mathrm{exc}}.  \label{DR-exchange-general}
\end{equation}%
Here $\chi _{\mathrm{cl}}$ is the classical electron susceptibility, given
by $\chi _{\mathrm{cl}}=(e^{2}/\varepsilon _{0}m)\int f_{0}d^{3}v/\left(
\omega -kv_{z}\right) ^{2}$, and $\chi _{\mathrm{exc}}$ is the exchange
correction to the susceptibility, given by 
\begin{equation}
\chi _{\mathrm{exc}}=\frac{\hbar ^{2}k\omega _{p}^{4}}{2\pi m^{2}v_{T}^{3}}%
\int du_{\perp }du_{z}dw_{z}\frac{1}{\left( \omega -kv_{T}w_{z}\right) ^{2}}%
\frac{u_{\perp }u_{z}}{u_{\perp }^{2}+u_{z}^{2}}\frac{%
(w_{z}-u_{z})(w_{z}+u_{z})}{\left[ \omega -kv_{T}(w_{z}-u_{z})\right] }\exp
\left( -u_{\perp }^{2}-u_{z}^{2}-w_{z}^{2}\right)   \label{exchange-suscept1}
\end{equation}%
where the velocity integrations (normalized against the thermal velocity) covers all of velocity space. In
general, the last integrals to find $\chi _{\mathrm{exc}}$ must be solved
numerically. However, the case of most interest is when  $kv_{T}/\omega \ll 1
$, such that the Landau damping is weak, in which case we can expand the
integrals in powers of $kv_{T}/\omega $. To leading order in $kv_{T}/\omega $%
, neglecting the pole contribution associated with Landau damping
altogether, Eq. (\ref{DR-exchange-general}) reduces to%
\begin{equation}
\omega ^{2}=\omega _{p}^{2}+3k^{2}v_{T}^{2}\left( 1-\frac{1}{90}\frac{\hbar
^{2}\omega _{p}^{2}}{m^{2}v_{T}^{4}}\right) .  \label{DR}
\end{equation}%
After correcting a slight numerical error in equation (49) of Ref.\ \cite{Roos} we note that the result there is exactly a factor two larger than our
result above. The difference is due to that Ref.\ \cite{Roos} does not take
into the spin part of the wavefunction. We note that it is the full many-body wavefunction that should be
anti-symmetric with respect to particle interchange, not just the spatial part. Ignoring this over-estimates the exchange correction by a factor of two.

While the scaling with temperature and density is the same as for many other
quantum phenomena (proportional to $H^{2}=\hbar ^{2}\omega
_{p}^{2}/m^{2}v_{T}^{4}$)$\,$, we note that the overall factor also contains
the small dimensionless number $1/90$ which appears for geometrical reasons.
Due to this small number, it makes sense to ignore exchange effects at the
same time as other quantum effects are kept. However, as we will see below,
there is no general principle guaranteeing the relative smallness of
exchange effects. Thus, whether or not exchange effects can be ignored for
other specific problems, as compared to particle dispersive effects, is an open question in general. The next sub-section will deal with ion-acoustic waves. However, before ending this sub-section, we note that the case of degenerate electrons require different integrals to be solved. We omit this case here, but point out that the exchange contribution to the Langmuir wave dispersion relation with a fully degenerate Fermi-Dirac background distribution can be found in Ref \cite{Egen-2015-second-exchange}.

\subsection{Low-frequency ion-acoustic waves}

The general procedure numbered 1-3 of the previous sub-section still applies for ion-acoustic waves. However, the concrete integrals that need to be solved depend on whether the electrons are degenerate or non-degenerate (assuming we limit ourselves to a thermodynamic background distribution). Also, the integration will be simplified in the quasi-neutral regime, that applies for frequencies $\omega\ll \omega_{pi}$. An additional thing to consider is the Landau damping, that often can be neglected for Langmuir waves, but generally tends to be significant for ion-acoustic waves. Thus we must use the Landau contours when evaluating the integrals, and keep track of both the real and imaginary parts. The case of cold classical ions and a Maxwell Boltzmann distribution for electrons was considered by Ref. \cite{Egen-2013-exchange}, and the dispersion relation in the quasi-neutral limit was found to be

\begin{equation}
\omega = kc_{s}\left( 1+ 0.8 \frac{\hbar ^{2}\omega _{pe}^{2}}{%
m^{2}v_{Te}^{4}}\right) -i\gamma _{\mathrm{cl}}\left( 1 - 3 \frac{\hbar
^{2}\omega _{pe}^{2}}{m^{2}v_{Te}^{4}}\right)  \label{Simplified}
\end{equation}%
where $c_{s}=\left( m_{e}/m_{i}\right) ^{1/2}v_{Te}$ is the classical
ion-acoustic velocity and we have introduced the classical electron Landau
damping, $\gamma_{\text{cl}} = k c_s \sqrt{\pi/8} \sqrt{m_e/m_i}$, in the
cold ion limit \cite{boyd}. The coefficients of the real and imaginary quantum terms ($0.8$ and $3$, respectively) are only approximate, as the final step involves a numerical integration. An important result from this calculation is that the exchange corrections now is proportional to a factor of the order unity times the quantum parameter $H^2$, in contrast to the previous case of Langmuir waves. As a consequence, when studying short scale dynamics, it is not a good approximation to keep the particle dispersive terms through the Wigner-Moyal equation, and simultaneously drop the exchange contribution. 

Next, we turn our attention to the same case as above, but with degenerate electrons.  In the quasi-neutral limit, the dispersion relation can then be computed as (see, e.g. Ref. \cite{Egen-2015-second-exchange} for details]
\begin{equation}
\omega ^{2}=\alpha ^{2}k^{2}v_{\text{F}}^{2}\left[ 1-\frac{\hbar ^{2}\omega
_{e}^{2}}{3m_{e}^{2}v_{\text{F}}^{4}}(14.9+7.11i)\right]
\label{Ion-acoustic-DR}
\end{equation}%
where $\alpha=\sqrt{m_e/3m_i}$. Apparently, the relative magnitude of the exchange terms are even larger than for the non-degenerate case, and, accordingly, independently of the background distribution, the omission of exchange terms in the quantum regime cannot be justified for ion-acoustic waves. 
\subsection{Exchange effects - final discussion}
\begin{figure*}
	\begin{subfigure}{0.47\textwidth}
		\includegraphics[width=\textwidth]{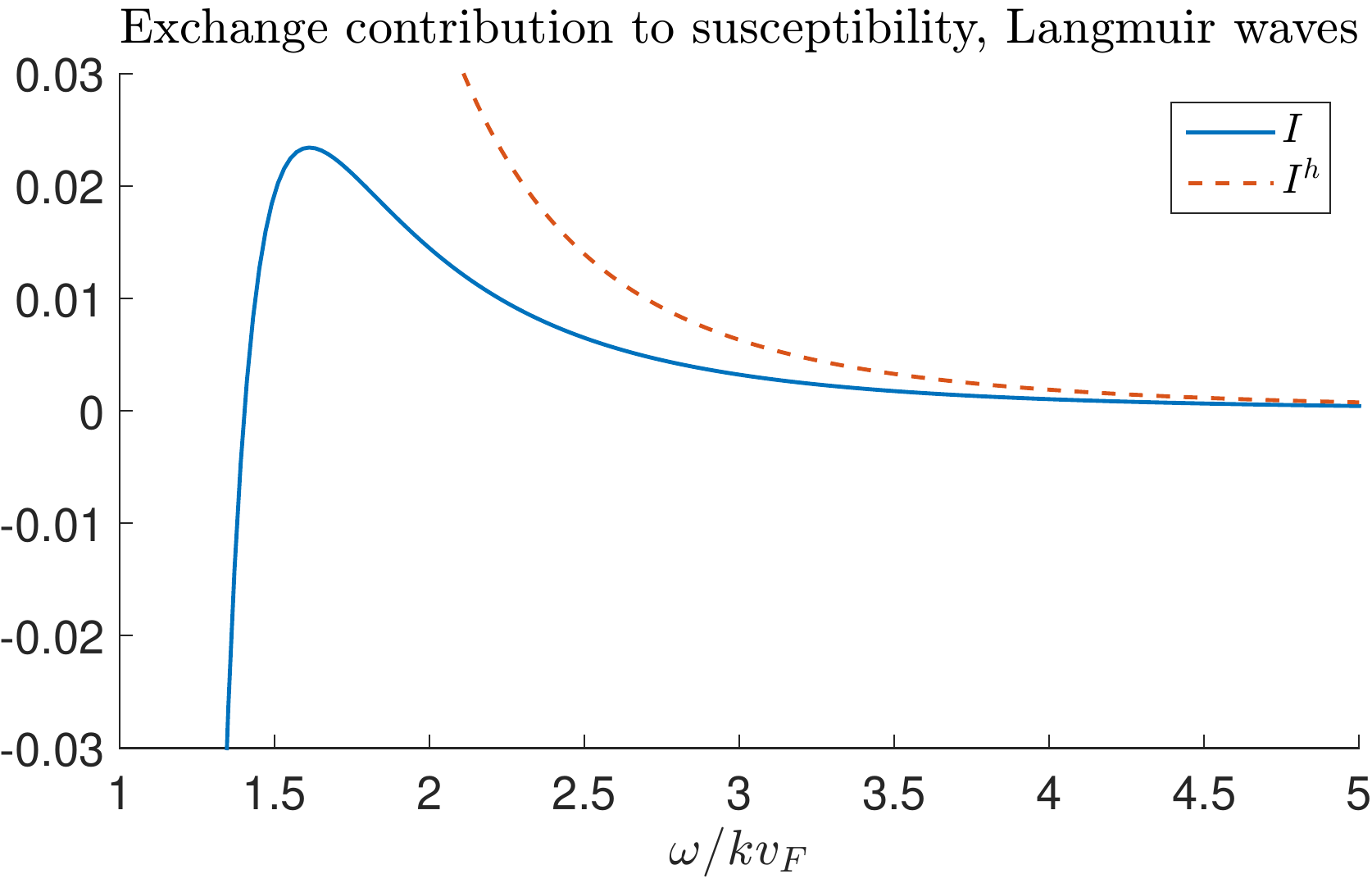}
		\caption{ \label{fig:langmuir-full}}
	\end{subfigure}
	\begin{subfigure}{0.47\textwidth}
		\includegraphics[width=\textwidth]{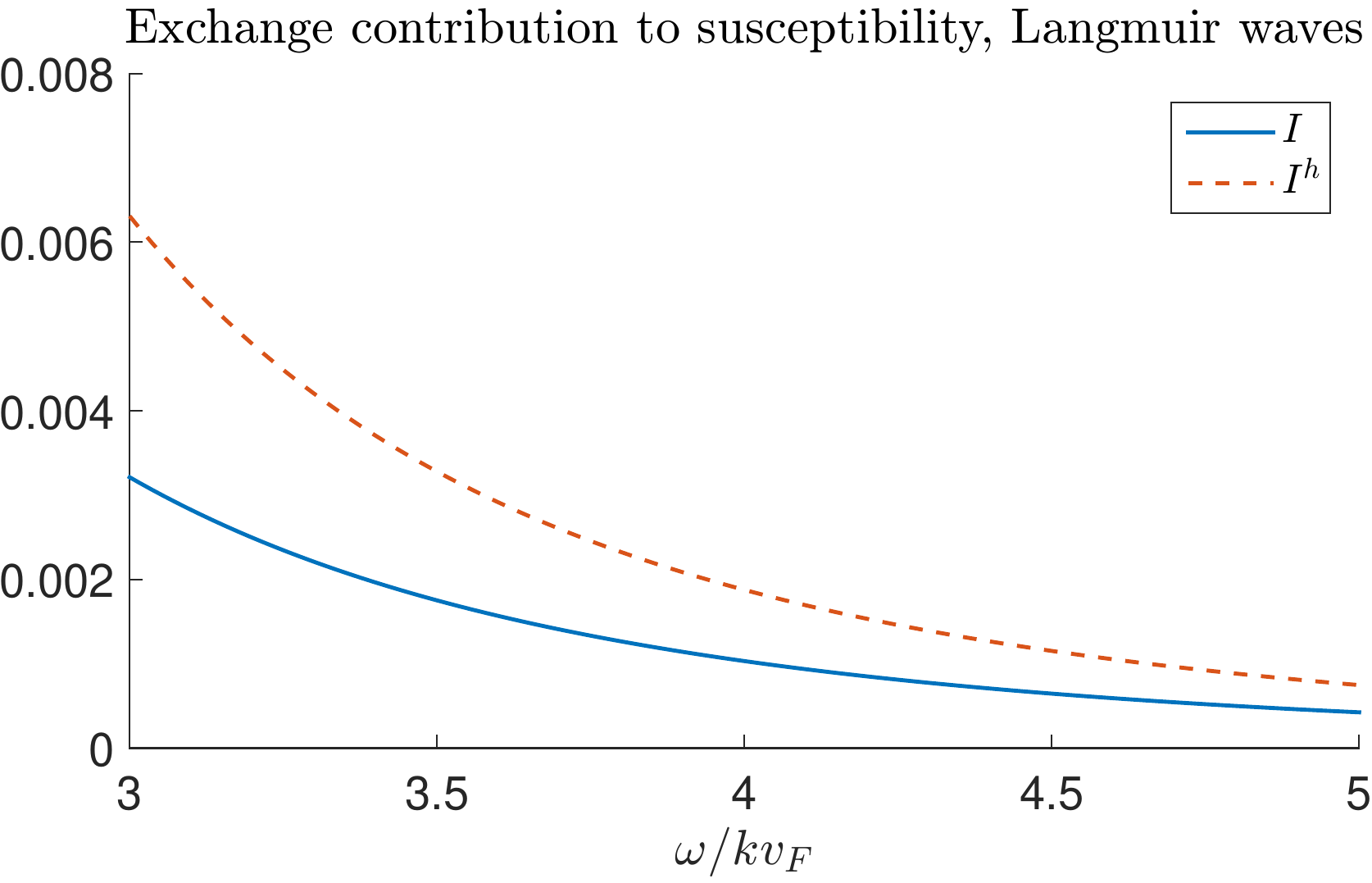}
		\caption{\label{fig:langmuir-zoomed} }		
	\end{subfigure}
	\caption{Exchange contribution to the susceptibility, kinetic (solid) and hydrodynamic (dashed), for Langmuir modes.
	\subref{fig:langmuir-full} Full range of $\omega/kv_F$
	\subref{fig:langmuir-zoomed} Detail of the high-frequency regime, showing different limits for $I$ (the normalized kinetic exchange contribution) and $I^h$ (the normalized hydrodynamic exchange contribution). Reproduced from Ref. \cite{Egen-2015-exchange}, with the permission of AIP Publishing.
\label{fig:langmuir}}
\end{figure*}

\begin{figure}
	\includegraphics[width=0.6\textwidth]{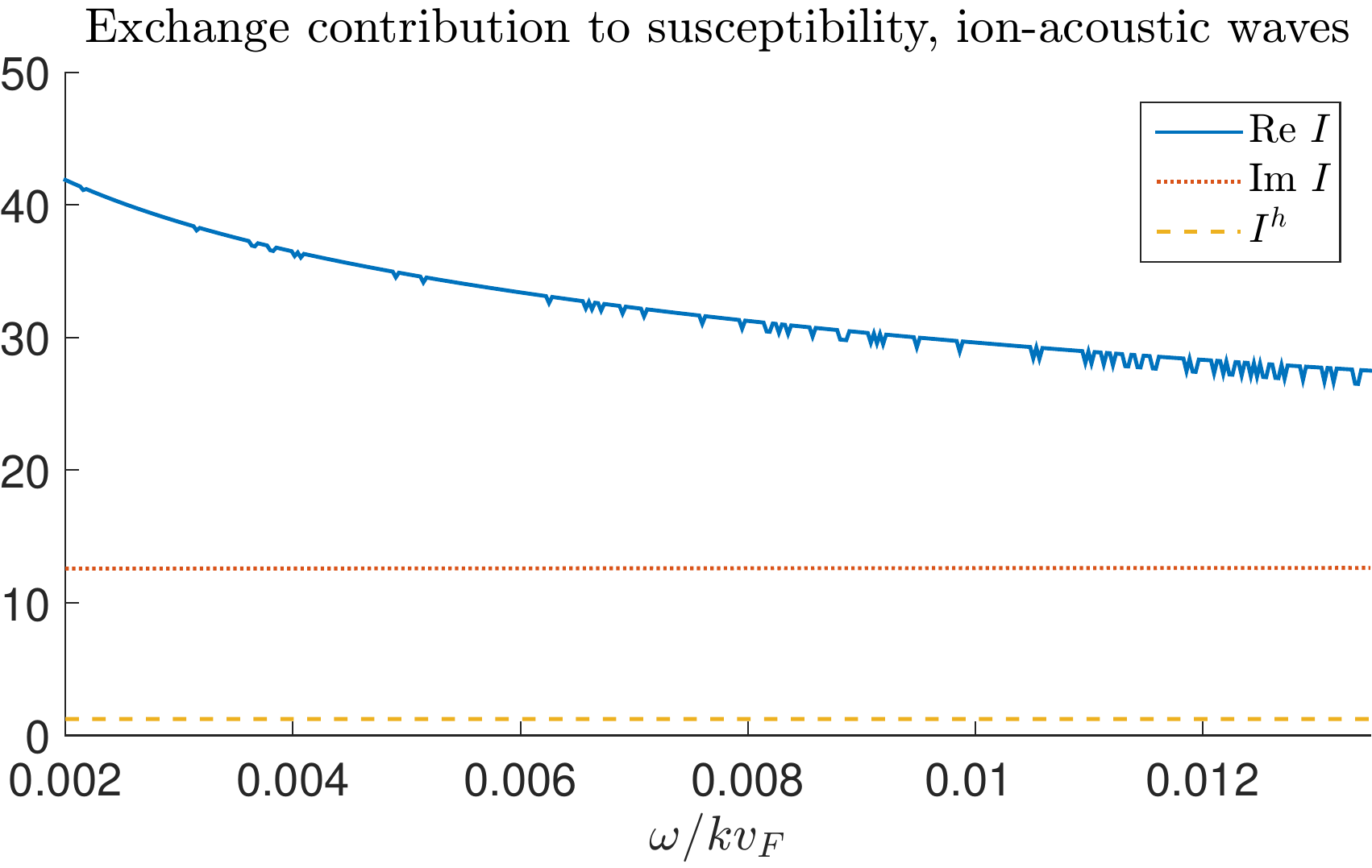}
	\caption{The exchange contribution to the susceptibility, kinetic (solid) and hydrodynamic (dashed), for ion-acoustic modes.
	Note that in the kinetic case, there is also an imaginary part (dotted).
	The irregularities in the real part of $I(\omega/kv_F)$ are due to the numerical resolution. Reproduced from Ref. \cite{Egen-2015-exchange}, with the permission of AIP Publishing.
	\label{fig:ionac}}
\end{figure} 
As illustrated above, due to the non-trivial momentum integrals, a first principal kinetic treatment of exchange effects is complicated. Going beyond the simplest cases, like the ones studied above, would require a major numerical effort. Even so, only perturbative treatments where exchange effects are small would be possible. However, the topic is an important one, as exchange effects are not generally small compared to other quantum effects. A tentative conclusion, suggested by the above findings, would be that the exchange contribution might be limited (to the extent that it could be negligible) for high-frequency phenomena, but not for low-frequency phenomena. However, more studies need to be done to put such a conclusion on a firm ground. 

An important generalization of the above treatment, is to cover also electromagnetic phenomena. Some key steps toward that goal was taken in Ref. \cite{Egen-2015-exchange}, although no concrete examples were worked out. An important issue to address in this case is the gauge invariance, as the exchange integrals tend to be dependent on the electromagnetic potentials, rather than the electromagnetic fields. 

In light of the challenging nature of the kinetic exchange contribution, it would be valuable to have less complex models, allowing for computational progress also beyond the perturbative regime. A quantum hydrodynamical model fulfilling this criterion has been put forward by Ref. \cite{Manfredi-exchange}. In addition to exchange effects, this model also covers the effect of electron correlations. Importantly, the hydrodynamical model is easy to use for practical calculation, as the exchange and correlation terms does not add much extra difficulty compared to pressure and particle dispersive terms. However, a drawback is that the model is based on time-independent density functional theory, and hence the applicability to dynamical phenomena is uncertain at best. A comparison of the hydrodynamical and kinetic models, to some extent covering regimes not presented above, has been made in Ref. \cite{Egen-2019-exchange}, and the numerical comparisons are shown in Figs 1 and 2, borrowed with permission from the original source.  The overall conclusion is that there is a reasonable agreement between the hydrodynamic and the kinetic models in the long wavelength high-frequency regime, but otherwise the hydrodynamical model tend to largely underestimate the importance of exchange effects.

\section{The Pauli Hamiltonian and Spin dynamics}
\label{section4}

Next we will generalize the treatment based on the Scr\"{o}dinger equation
to include the spin dynamics. In principle the formalism is similar to that
of section II. In particular the higher order terms in the BBGKY-hierarchy \cite{BBGKY} are
dropped, and thus the evolution equation will be derived from the single
particle density matrix, using the von Neumann equation, followed by a
Wigner-transform. Nevertheless, there is an important difference compared to
the previous case, as the density matrix now depend on the spin state. As a
result, the single component in the Scr\"{o}dinger case is replaced by a two
by two matrix, where the different components represent different spin
states.  The changes needed to obtain the kinetic evolution equation are
outlined in subsection A, and two different but equivalent systems are presented.   Then, in subsection B, we use the model to derive
the linear conductivity tensor in a magnetized plasma, generalizing previous
results to cover the case of an an-isotropic background distribution.   

\subsection{Derivation of evolution equations}

Including the spin dynamics, in the lowest (non-relativisitic)
approximation, we replace the Schr\"{o}dinger Hamiltonian of Section II with
the Pauli Hamiltonian%
\begin{equation}
\hat{H}=mc^{2}+q\phi +\frac{1}{2m}\left( \hat{\mathbf{p}}-q\mathbf{%
A}\right) ^{2}-\frac{q\hbar }{2m}\bm{\sigma}\cdot \mathbf{B}
\label{eq:Ham-Pauli}
\end{equation}%
where $\bm{\sigma}$ here denotes a vector with the $2\times 2$ Pauli
matrices as components.  Following Ref. \cite{Zamanian 2010}, we are able
to construct a gauge invariant scalar kinetic theory using a density matrix
description for a spin-1/2 particle.

The basis states are $|\mathbf{x},\alpha \rangle =|\mathbf{x}\rangle \otimes
|\alpha \rangle $, where $|\mathbf{x}\rangle $ is a state with position $%
\mathbf{x}$ and $|\alpha \rangle $ is the state with spin-up $\alpha =1$ or
spin-down $\alpha =2$. As a starting point for the derivation, we use the
spinor state $\psi \left( \mathbf{x},\alpha ,t\right) =\langle \mathbf{x}%
,\alpha |\psi \rangle $ which fulfill the dynamical equation $i\hbar
\partial _{t}\psi \left( \mathbf{x},\alpha ,t\right) =\hat{H}\psi \left( 
\mathbf{x},\alpha ,t\right) $, with the Hamiltonian (\ref{eq:Ham-Pauli}). The
density matrix is now given by 
\begin{equation}
\rho _{\alpha \beta }\left( \mathbf{x},\mathbf{y},t\right) =\langle \mathbf{x%
},\alpha |\hat{\rho}|\mathbf{y},\beta \rangle =\sum_{i}p_{i}\psi _{i}\left( 
\mathbf{x},\alpha ,t\right) \psi _{i}^{\dag }\left( \mathbf{y},\beta
,t\right) \,,
\end{equation}%
where, as before, $p_{i}$ is the probability to have a state ${\psi }_{i}$,
but we have the additional dependence on spin state. Similarly as before,
the von Neumann equation applies, i.e. the evolution equation for the
density matrix is  
\begin{equation}
i\hbar \frac{\partial \hat{\rho}_{\alpha \beta }}{\partial t}=\left[ \hat{H},%
\hat{\rho}_{\alpha \beta }\right] \,.  \label{vonneumann}
\end{equation}%
Once the density matrix has been defined, we can define the
Wigner-Stratonovich transform \cite{strato} as 
\begin{equation}
W_{\alpha \beta }(\mathbf{x},\mathbf{p},t)=\int \frac{d^{3}z}{(2\pi \hbar
)^{3}}\exp \left[ -\frac{i}{\hbar }\mathbf{z}\cdot \Phi \right] \rho
_{\alpha \beta }\left( \mathbf{x}+\frac{\mathbf{z}}{2};\mathbf{x}-\frac{%
\mathbf{z}}{2},t\right) \,,  \label{wstrans}
\end{equation}%
where the phase 
\begin{equation}
\Phi =\mathbf{p}-q\int_{-1/2}^{1/2}d\eta \mathbf{A}\left( \mathbf{x%
}+\eta \mathbf{z},t\right) \,
\end{equation}%
is used to ensure gauge invariance of the resulting distribution function.
This is the same transformation as in Eq.~\eqref{wigstrat} with the modification 
that it must to be taken separately for each component of the 2-by-2 density matrix. 

In principle, Eq. (\ref{vonneumann}) together with Eq. (\ref{wstrans}) gives
us a  kinetic evolution equation, and we could be content with this.
However, the individual components of $W_{\alpha \beta }$ have no clear
physical interpretation, and it is desirable to construct functions that can
be understood more intuitively. Two ways to do this has been presented in
the literature, that we will descibe below. 

The first way is to make a spin transform (or Q-transform, as it has also
been called \cite{Q-transform}), which creates a single scalar function $f$ from the
Wigner matrix $W_{\alpha \beta }$. The attractive feature here is that the
charge and current sources in Maxwell's equation can be calculated from a
single scalar function $f$, which plays a role much like the distribution
function of classical physics. However, there is a prize to pay, since the
transform extends the classical phase space to include an extra \textit{%
independent} variable, namely the spin. Thus our scalar function has the
functional dependence $f=f(\mathbf{x},\mathbf{p},\mathbf{s},t)$.       

Following this approach, thoroughly discussed in Ref.\ \cite{Zamanian 2010}, we
here define a scalar distribution function $f(\mathbf{x},\mathbf{p},\mathbf{s%
},t)$ in the extended phase-space as  
\begin{align}
f(\mathbf{x},\mathbf{p},\mathbf{s},t)& =\frac{1}{4\pi }\sum_{\alpha ,\beta
=1}^{2}\left( 1+\mathbf{s}\cdot \bm\sigma \right) _{\alpha \beta }W_{\beta
\alpha }(\mathbf{x},\mathbf{p},t)  \notag \\
& =\frac{1}{4\pi }\mathrm{tr}(1+\mathbf{s}\cdot \bm\sigma )W(\mathbf{x},%
\mathbf{p},t),  \label{Q-trans}
\end{align}%
where $\mathbf{s}$ is a vector of unit length and $\mathrm{tr}$ denotes the
trace over the spin indices. Integrating over momentum,   
\begin{equation}
f(\mathbf{x},\mathbf{s},t)=\int d^{3}pf(\mathbf{x},\mathbf{p},\mathbf{s}%
,t)\,,
\end{equation}%
we find that the reduced distribution function $f(\mathbf{x},\mathbf{s},t)$
gives the probability to find the particle at position $\mathbf{x}$ with
spin-up in the direction of $\mathbf{s}$. Similarly, defining 
\begin{equation}
f(\mathbf{p},\mathbf{s},t)=\int d^{3}xf(\mathbf{x},\mathbf{p},\mathbf{s}%
,t)\,,
\end{equation}%
we find that $f(\mathbf{p},\mathbf{s},t)$ gives the probability to find the
particle with momentum $\mathbf{p}$ with spin-up in the direction of $%
\mathbf{s}$.  Moreover, we note that the expectation value for the spin
polarization density is now given by 
\begin{equation}
\left\langle \bm\sigma \right\rangle (\mathbf{x},t)=\mathrm{tr}[\bm\sigma
\rho (\mathbf{x},\mathbf{y},t)]=3\int d^{3}pd^{2}sf(\mathbf{x},\mathbf{p},%
\mathbf{s},t)\mathbf{s},  \label{spint-calc}
\end{equation}%
where we stress the need for the factor 3. This follows from the form of the
transformation \eqref{Q-trans} and is needed to compensate for the quantum
mechanical smearing of the distribution function in spin space. Eqs.~(\ref{vonneumann}), (\ref{wstrans}) and (\ref{Q-trans}) determines the evolution
equation for $f(\mathbf{x},\mathbf{p},\mathbf{s},t)$. We refer to Ref. \cite{Zamanian 2010} for a more thorough description of the technical details, and move on to the final result for the evolution equation, which is given by
\begin{eqnarray}
	\frac{\partial f}{\partial t} + (\mathbf v + \Delta \tilde{\mathbf v} ) \cdot \nabla_\mathbf{x} f + 
	\frac{q}{m} \left[ (\mathbf v + \Delta \tilde{\mathbf v} ) \times \tilde{\mathbf B} + 
	\tilde{\mathbf E} \right] \cdot \nabla_\mathbf{v} f + \nonumber\\ 
	\frac{\mu}{m} \nabla_\mathbf{x} [ (\hat{\mathbf s} + \nabla_{\hat s} ) \cdot \tilde{\mathbf B} ] 
	\cdot \nabla_\mathbf{v} f 
	+ \frac{2\mu}{\hbar} \left[ \hat{\mathbf s} \times
	\left(  \tilde{\mathbf B} + \Delta \tilde{\mathbf B} \right) \right] \cdot \nabla_{\hat{\mathbf{s}}} f = 0, 
	\label{82}
\end{eqnarray}
where, as in Eq.~\eqref{gaugeevolu}, we have defined 
\begin{eqnarray}
	&&\!\!\!\!\!\!\!\!\!\!\!\!\!\!\!\!\!\!
	\tilde{\mathbf E} 
	= \int_{-1/2}^{1/2} d\tau \mathbf E \left(\mathbf x + \frac{i\hbar\tau}{m} 
	\nabla_\mathbf{v} \right) 
	= \mathbf{E}(\mathbf{x}) \int_{-1/2}^{1/2}d\tau\, \cos\left(  \frac{\tau\hbar}{m}\stackrel{\leftarrow}{\nabla}_\mathbf{x}\cdot
	 \stackrel{\rightarrow}{\nabla}_\mathbf{v}\right) \label{Etilde} 
\\ &&\!\!\!\!\!\!\!\!\!\!\!\!\!\!\!\!\!\!
	\tilde{\mathbf B}  
	= \int_{-1/2}^{1/2} d\tau \mathbf B \left(\mathbf x + \frac{i\hbar\tau}{m} 
	\nabla_\mathbf{v} \right) 
	= \mathbf{B}(\mathbf{x})\int_{-1/2}^{1/2}d\tau\, \cos\left(  \frac{\tau\hbar}{m}\stackrel{\leftarrow}{\nabla}_\mathbf{x}\cdot
	 \stackrel{\rightarrow}{\nabla}_\mathbf{v}\right)  
\\ &&\!\!\!\!\!\!\!\!\!\!\!\!\!\!\!\!\!\!
	 	\Delta \tilde{\mathbf v} 
		= - \frac{iq\hbar}{m^2} \int_{-1/2}^{1/2} 
	d\tau\, \tau \mathbf B \left( \mathbf x + \frac{i \hbar \tau}{m} \nabla_\mathbf{v}  \right) \times \nabla_\mathbf{v} 
	= 
	\frac{q\hbar}{m^2}\left[ \mathbf B(\mathbf x) \int_{-1/2}^{1/2} 
	d\tau\, \tau \sin\left( \frac{\tau\hbar}{m}\stackrel{\leftarrow}{\nabla}_\mathbf{x}\cdot
	 \stackrel{\rightarrow}{\nabla}_\mathbf{v} \right)\right] \times \stackrel{\rightarrow}{\nabla}_\mathbf{v} 
	 \label{dvtilde}
\\ &&\!\!\!\!\!\!\!\!\!\!\!\!\!\!\!\!\!\!
	\Delta \tilde{\mathbf B} 
	= - \frac{i\hbar}{m} \int_{-1/2}^{1/2} d\tau\, \tau 
	\mathbf B \left( \mathbf x + \frac{i\hbar\tau}{m}\nabla_\mathbf{v} \right)  
	 \stackrel{\leftarrow}{\nabla}_\mathbf{x} \cdot \stackrel{\rightarrow}{\nabla}_\mathbf{v} 
	 =
	 \frac{\hbar}{m}\mathbf{B}(\mathbf{x})\int_{-1/2}^{1/2}d\tau\, \tau\sin\left( 
	 	\frac{\tau\hbar}{m}\stackrel{\leftarrow}{\nabla}_\mathbf{x} \cdot \stackrel{\rightarrow}{\nabla}_\mathbf{v}
	 \right)\stackrel{\leftarrow}{\nabla}_\mathbf{x} \cdot \stackrel{\rightarrow}{\nabla}_\mathbf{v} . \label{dBtilde}
\end{eqnarray}	
In the long scale limit, for spatial variations much longer than the characteristic de Broglie length, the local approximations $\tilde{\mathbf E}\approx {\mathbf E}$, etc, apply, and the term $\propto \Delta\tilde{\mathbf v}$ can be dropped altogether. As a result, the equation accounting for spin dynamics, but dropping short scale physics, is given by
\begin{equation}
	\frac{\partial f}{\partial t} + \mathbf v \cdot \nabla_\mathbf{x} f + 
	\left[ 
		\frac{q}{m}(\mathbf{E} + \mathbf v\times \mathbf{B})\cdot \nabla_\mathbf{v} + 
		\frac{\mu_B}{m} \nabla_\mathbf{x} [ (\hat{\mathbf s} + \nabla_{\hat{\mathbf{s}}} ) \cdot \mathbf{B} ] \cdot \nabla_\mathbf{v} + \frac{2\mu_B}{\hbar}(\hat{\mathbf{s}}\times\mathbf{B})\cdot\nabla_{\hat{\mathbf{s}}}
	\right] f=0
	\label{paulilong}
	\end{equation}
Together with Maxwell's equations, Eq. (\ref{82}) (or Eq. (\ref{paulilong}) in the long scale limit)) provide a closed description for the spin dynamics, where the electron charge density is given by \begin{eqnarray}
\rho = q_{e}\int fd^{2}sd^{3}v
\end{eqnarray}
and the current density is given by \begin{eqnarray}
\mathbf{J} &=&\mathbf{J}_{f}+\mathbf{J}_{M}  \notag  \label{J} \\
&=&\mathbf{J}_{f}+\nabla _{\mathbf{x}}\times \mathbf{M}  \notag \\
&=&q_{e}\int \mathbf{v}fd^{2}sd^{3}v+\nabla _{\mathbf{x}}\times \left( \mu
_{e}\int 3\hat{\mathbf{s}}fd^{2}sd^{3}v\right) .
\label{paulicurrent}
\end{eqnarray}%
Here the two-dimensional spin integration is made over the Bloch sphere (naturally represented in spherical spin coordinates), and ${\hat{s}}$ is the spin unit vector. As done above, the current density is naturally divided into its free contribution $\mathbf{J}_{f}$ and the contribution $\mathbf{J}_{M}$ due to the spin magnetization. The physics of the evolution equation can be understood as follows. The basic effect of short scale particle dispersion is captured in the non-local variables defined in Eqs. (\ref{Etilde})-(\ref{dBtilde}), and the same physics is present already without the spin effects of the Pauli Hamiltonian, as described already in section II. The effects due to the spin that are genuinely new all survive in the long scale limit, as seen in Eq. (\ref{paulilong}). The first terms of the equation are familiar from classical physics, then we have the effects of the magnetic dipole force, and the last term is the spin precession. To a large degree, Eq. (\ref{paulilong}) agrees with a semi-classical kinetic theory \cite{Brodin2008}. The main difference is that the magnetic dipole force has a slightly more complicated dependence $\propto \nabla_x [ (\hat{\mathbf s} + \nabla_{\hat s} ) \cdot \mathbf{B} ] 		\cdot \nabla_v f$ rather than the simpler semi-classical expression $\propto \nabla_x [ (\hat{\mathbf s}) \cdot \mathbf{B} ] 		\cdot \nabla_v f$. The reason for this difference is discussed in Ref. \cite{Zamanian 2010}. 

In the next sub-section, we will demonstrate how to handle the extra spin-dependence in practical calculations. However, it is not necessary to use the Q-transform and introduce spin as an extra independent variable. Instead, we can proceed as Ref. \cite{Manfredi-spin1} and split the Wigner matrix $W_{ab}$ into a vector part and a scalar part. 

Specifically, we can split the Wigner matrix $W_{ab}$ into the scalar, defined as 
\begin{equation}
f_{0}=\mathrm{tr} W_{ab} = W_{11} + W_{22}  \label{scalar}
\end{equation}%
and the vector $\mathbf{f}$ defined as  
\begin{equation}
\mathbf{f=}\left( \bm{\sigma}W_{ab}\right) .  \label{vector}
\end{equation}%
With these definitions, the Wigner matrix can be reconstructed as 
\begin{equation}
W_{ab}=\frac{1}{2}\sigma _{0}f_{0}+\frac{1}{2}\mathbf{f\cdot \bm{\sigma}}
\label{reconstruct}
\end{equation}%
where the Pauli spin matrices building up the vector $\bm{\sigma }$ has
been complemented with the unit matrix $\sigma _{0}.$ With the aid of 
(\ref{scalar})-(\ref{reconstruct}), the equation for the Wigner matrix can
be rewritten in terms of $f_{0}$ and $\mathbf{f}$, with the result 
\begin{equation}
\frac{\partial f_{0}}{\partial t}+(\mathbf{v}+\Delta \tilde{\mathbf{v}}%
)\cdot \nabla_{\mathbf{x}}f+\frac{q}{m}\left[ \tilde{\mathbf{E}}+(\mathbf{v}+\Delta 
\tilde{\mathbf{v}})\times \tilde{\mathbf{B}}\right] \cdot \nabla_{\mathbf{v}}f_{0}+%
\frac{\mu }{m}\nabla _{x}[\tilde{\mathbf{B}}_{i}]\cdot \nabla_{\mathbf{v}}f_{i}=0,
\label{Manfred-scalar}
\end{equation}%
and%
\begin{equation}
\frac{\partial f_{i}}{\partial t}+(\mathbf{v}+\Delta \tilde{\mathbf{v}}%
)\cdot \nabla_{\mathbf{x}}f_{i}+\frac{q}{m}\left[ \tilde{\mathbf{E}}+(\mathbf{v}%
+\Delta \tilde{\mathbf{v}})\times \tilde{\mathbf{B}}\right] \cdot \nabla_{\mathbf{v}}
f_{0}+\frac{\mu }{m}\nabla_{\mathbf{x}}[\tilde{\mathbf{B}}_{i}]\cdot \nabla_{\mathbf{v}}
f_{0}=0,  \label{Manfred-vector}
\end{equation}%
where the definitions for $\tilde{\mathbf{E}}$, $\Delta \tilde{\mathbf{v}}$, and 
$\tilde{\mathbf{B}}$ are the same as in (\ref{Etilde})-(\ref{dvtilde}). As expected, Eqs.~(\ref{Manfred-scalar}) and (\ref{Manfred-vector})
are completely equivalent to (\ref{82}). The scalar function $f_{0}$ captures the
phase space density, that is the charge density is computed as 
\begin{equation}
\rho = q_{e}\int f_{0}d^{3}v  \label{manfredi-charge}
\end{equation}%
whereas the vector $\mathbf{f}$ gives the spin density (which in the
approximation of the Pauli Hamiltonian used here coincides with the
magnetization). Thus the magnetization current is given by 
\begin{equation}
\mathbf{j}_{m}=\nabla_{\mathbf{x}}\times \mathbf{M}=\mu \nabla _{x}\times \left[
\int \mathbf{f}d^{3}v\right]   \label{Magnetization-manf}
\end{equation}%
and the full current density to be used in Ampere's law therefore is 
\begin{eqnarray}
\mathbf{j} &=&\mathbf{j}_{m}+\mathbf{j}_{f}  \notag \\
&=&\nabla_{\mathbf{x}} \times \mathbf{M}  + q_{e}\int \mathbf{v}f_{0}d^{3}v
\label{Manf-current}
\end{eqnarray}%
Note that In the long-scale limit, the approximations  $\tilde{\mathbf{E}}%
\rightarrow \mathbf{E}$, etc., applies, in which case the long-scale version
of (\ref{Manfred-scalar}) and (\ref{Manfred-vector}) agrees with Eq. (\ref{paulilong})). This version has been used by Ref. \cite{Manfredi-spin1} to
derive fluid equation, with the aid of moment expansions. For more
applications of Eqs. (\ref{Manfred-scalar}) and (\ref{Manfred-vector}), see e.g. Refs. \cite{Manfredi-2019,Manfredi-solid}. 

After presenting two different (but equivalent) systems derived from the
Wigner matrix, one can ask what has been gained. After all, all the physics
is already contained in the equations for $W_{ab}$. However, the coupled
equations for the components of $W_{ab}$ does not provide much help of
guiding the physical intuition. The individual components do not carry physical meaning themselves, only the complete object does. By contrast, the
scalar object $f(\mathbf{r},\mathbf{v},\mathbf{s},t)$ can be applied much
like a classical distribution function, but in an extended phase space.
Similarly, $f_{0}$ can be viewed much like a classical distribution
function, but in this case averaged over the spin state, whereas the spin
properties is captured in the vector $\mathbf{f}$. As a consequence, the
quantities of the two theories can be related by the relations $f_{0}(%
\mathbf{r},\mathbf{v},t)=\int f(\mathbf{r},\mathbf{v},\hat{\mathbf{s}}%
,t)d^{2}s$ and $\mathbf{f(\mathbf{r},\mathbf{v},t)=}\int \mathbf{s}f(\mathbf{%
r},\mathbf{v},\hat{\mathbf{s}},t)d^{2}s$. 

Each of the two formulations has its own advantages. Firstly,  the latter
formulation (Eqs. (\ref{Manfred-scalar}) and (\ref{Manfred-vector})) give somewhat shorter equations. There is no
explicit spin precession term, and the magnetic dipole force is now a bit
simpler (one does not need the operator $(\hat{\mathbf{s}}+\nabla _{\hat{%
\mathbf{s}}})$). As a result these equations constitute a rather direct
extension of the Vlasov equation. More importantly, not having the extra
independent variable $\hat{\mathbf{s}}$, makes the latter equations more
suitable for a direct numerical approach. On the other hand, Eq. (\ref{82}) is
attractive when aiming for analytical solutions. Since only a scalar
function is involved, and the structure of the extra terms are similar to
the classical ones, most analytical approaches developed for the Vlasov equation can
be adopted directly, with the simple addition of an extra integration over
spin space. Moreover, in the case of the long scale version, Eq. (\ref{paulilong}),
replacing the magnetic dipole force with its semi-classical correspondence,
allows the equation to be written in a form consistent with classical
PIC-schemes (see Ref. \cite{PIC-manfredi}), in which case an efficient and attractive
numerical scheme is possible.  

\subsection{The linear conductivity tensor}

We will here limit ourselves to the case of long spatial scale-length (much
longer than the characteristic de Broglie length), in which case the local
approximations $\mathbf{\tilde{E}\rightarrow E}$, $\mathbf{\tilde{B}%
\rightarrow B}$, etc., applies. Thus the evolution equation for  $f(\mathbf{
r,v,}\hat{\mathbf{s}},t)$ is given by Eq. (\ref{paulilong}), and the current density to be used in Ampere's law is given by (\ref{paulicurrent}). 

In this section we only consider the electron contribution to the current
density, as the classical ion-contribution can be found in the textbook 
literature,  see e.g.\ Ref.\ \cite{Swanson}. found.  Below we will derive
the linear conductivity tensor $\sigma _{ij}$ in a homogeneous magnetized
plasma, defined as $J_{i}=\sigma _{ij}E_{j}$, for such a system, where $%
\sigma _{ij}$ contains both the free and magnetization current densities.
With the conductivity tensor known, it is straightforward to find the
dispersion relations for arbitrary wave modes. 

We start by linearizing the kinetic Eq.\ (\ref{paulilong}) according to 
$f=f_{0}+f_{1}$ and $\mathbf{B}=\mathbf{B}_{0}+\mathbf{B}_{1}$, where the
subscript $0$ denotes an unperturbed quantity and the subscript $1$ denotes
a perturbation, and we take $\mathbf{B}_{0}=B_{0}\hat{\mathbf{z}}$. Before
proceeding, let us point out a few quantum effects that may be contained
already in the unperturbed distribution function:

\begin{enumerate}
\item \textit{Fermi-Dirac statistics:} This effect is well-known. For $\hbar
^{2}n_{0}^{2/3}/m_{e}k_{B}T$ much larger than unity, we will have almost
complete degeneracy, whereas if the parameter is much smaller than unity, the
thermodynamic background distribution can be approximated by a Maxwellian. 

\item \textit{Landau-quantization:} The quantization of perpendicular energy
states becomes important in the regime of very strong magnetic fields, or
very low temperatures, when $\hbar \left\vert \omega _{ce}\right\vert
/k_{B}T\rightarrow 1$, where $k_{B}$ is Boltzmann's constant and $\omega
_{ce}=-eB_{0}/m_e$ is the electron cyclotron frequency.

\item \textit{Spin-splitting: }The two spin states, up- and down relative to
the magnetic field, have different probability distributions in spin space.
As a result, the general time-independent distribution function can be
written as $f_{0}=f_{0+}+f_{0-}$ with $f_{0\pm }=(1/4\pi )F_{0\pm }(\mathbf{v%
})(1\pm \cos \theta _{s})$, where for a time-independent distribution
function $F_{0\pm }$ can be arbitrary functions of $(v_{\perp },v_{z})$, and 
$F_{0\pm }$ is normalized such that $\int F_{0\pm }d^{3}v=n_{0\pm }$ with $%
n_{0\pm }$ being the number densities of the spin up/down states
respectively. The positive spin state here means that the spin points in the
direction parallel to the magnetic field, which means that the magnetic
moment points in the opposite direction. Note that with this definition, the
lower energy state is the spin state with negative index, i.e. $n_{0-}>n_{0+}
$ in case the background distribution $f_{0}$ describes a thermodynamic
equilibrium.
\end{enumerate}

For a full quantum mechanical expression of the thermodynamic equilibrium
distribution, see Ref.\ \cite{Zamanian 2010}. Next, we follow Ref. \cite{Lundin2010} but with a slight generalization,
allowing for a background distribution that is not necessarily isotropic,
i.e. we also cover wave modes that are subject to Weibel type of
instabilities. After linearization, Eq.\ (\ref{paulilong}) is written as 
\begin{eqnarray}
&&\frac{\partial f_{1}}{\partial t}+\mathbf{v}\cdot \nabla _{\mathbf{x}%
}f_{1}+\frac{q_{e}}{m_{e}}(\mathbf{v}\times \mathbf{B}_{0})\cdot \nabla _{%
\mathbf{v}}f_{1}+\frac{2\mu _{e}}{\hbar }(\hat{\mathbf{s}}\times \mathbf{B}%
_{0})\cdot \nabla _{\hat{\mathbf{s}}}f_{1}  \notag  \label{linVlasov} \\
&=&-\left[ \frac{q_{e}}{m_{e}}(\mathbf{E}+\mathbf{v}\times \mathbf{B}_{1})+%
\frac{\mu _{e}}{m_{e}}\nabla _{\mathbf{x}}\left( \hat{\mathbf{s}}\cdot 
\mathbf{B}_{1}+\mathbf{B}_{1}\cdot \nabla _{\hat{\mathbf{s}}}\right) \right]
\cdot \nabla _{\mathbf{v}}f_{0}-\frac{2\mu _{e}}{\hbar }(\hat{\mathbf{s}}%
\times \mathbf{B}_{1})\cdot \nabla _{\hat{\mathbf{s}}}f_{0}.
\end{eqnarray}
In order to proceed, we make a plane wave ansatz of the perturbed parameters according to $%
f_{1}=\tilde{f}_{1}\exp [i(\mathbf{k}\cdot \mathbf{x}-\omega t)]$, etc.
Without loss of generality, we define the wavevector as $\mathbf{k}=k_{\bot }%
\hat{\mathbf{x}}+k_{z}\hat{\mathbf{z}}$. We also choose to express the
velocity in cylindrical coordinates $(v_{\perp },\varphi _{v},v_{z})$ such
that $d^{3}v=v_{\bot }dv_{\bot }d\varphi _{v}dv_{z}$, and expand $f_{1}$ in
eigenfunctions to the operator of the right hand side 
\begin{equation}
\tilde{f}_{1}=\sum_{a=-\infty }^{\infty }\sum_{b=-\infty }^{\infty
}g_{ab}(v_{\perp },v_{z},\theta _{s})\psi _{a}(\varphi _{v},v_{\perp })\frac{%
1}{\sqrt{2\pi }}\exp (-ib\varphi _{s}),  \label{f1_expansion}
\end{equation}%
where 
\begin{eqnarray}
\psi _{a}(\varphi _{v},v_{\perp }) &=&\frac{1}{\sqrt{2\pi }}\exp
[-i(a\varphi _{v}-k_{\perp }v_{\perp }\sin \varphi _{v}/\omega _{ce})] 
\notag  \label{psi} \\
&=&\frac{1}{\sqrt{2\pi }}\sum_{l=-\infty }^{\infty }\mathcal{J}_{l}\left( 
\frac{k_{\perp }v_{\perp }}{\omega _{ce}}\right) \exp [i(l-a)\varphi _{v}],
\end{eqnarray}%
and $\mathcal{J}_{l}(x)$ is a Bessel function of the first kind. We may
then note the following simplifying relations; 
\begin{eqnarray}
\frac{q_{e}}{m_{e}}(\mathbf{v}\times \mathbf{B}_{0})\cdot \nabla _{\mathbf{v}%
}f_{1} &=&-\omega _{ce}\frac{\partial f_{1}}{\partial \varphi _{v}},  \notag
\\
\frac{2\mu _{e}}{\hbar }(\hat{\mathbf{s}}\times \mathbf{B}_{0})\cdot \nabla
_{\hat{\mathbf{s}}}f_{1} &=&-\omega _{cg}\frac{\partial f_{1}}{\partial
\varphi _{s}},
\end{eqnarray}%
where $\omega _{cg}=2\mu _{e}B_{0}/\hbar $ is the spin precession frequency.
Using the eigenfunction expansion of $\tilde{f}_{1}$ (Eq.\ (\ref%
{f1_expansion})) in the linearized Vlasov equation (\ref{linVlasov}),
multiplying the resulting equation with $\psi _{a}^{\ast }e^{ib\varphi _{s}}/%
\sqrt{2\pi }$ (where the star denotes complex conjugate) and integrating
over $\varphi _{v}$ and $\varphi _{s}$, we find the equation 
\begin{equation}
i(\omega -k_{z}v_{z}-a\omega _{ce}-b\omega _{cg})g_{ab}=I_{ab}(v_{\perp
},v_{z},\varphi _{s}),  \label{gab}
\end{equation}%
where 
\begin{multline}
I_{ab}=\int_{0}^{2\pi }\int_{0}^{2\pi }d\varphi _{v}d\varphi _{s}\left[
\left( \frac{q_{e}}{m_{e}}\left( \tilde{\mathbf{E}}+\mathbf{v}\times \tilde{%
\mathbf{B}}_{1}\right) +\frac{\mu _{e}}{m_{e}}\nabla _{\mathbf{x}}\left( 
\hat{\mathbf{s}}\cdot \tilde{\mathbf{B}}_{1}+\tilde{\mathbf{B}}_{1}\cdot
\nabla _{\hat{\mathbf{s}}}\right) \right) \cdot \nabla _{\mathbf{v}}f_{0}+%
\frac{2\mu _{e}}{\hbar }(\hat{\mathbf{s}}\times \tilde{\mathbf{B}}_{1})\cdot
\nabla _{\hat{\mathbf{s}}}f_{0}\right] 
\\ 
\times \psi _{a}^{\ast }\frac{1}{\sqrt{2\pi }%
}\exp (ib\varphi _{s}).  \label{Iab}
\end{multline}%
The result coincides with Ref. \cite{Lundin2010}, except that in our case $%
f_{0}(v_{\bot },v_{z})$ is an arbitrary function, which means that the term
in the integral $\propto \mathbf{v}\times \tilde{\mathbf{B}}_{1}$ survives
the integration $\int_{0}^{2\pi }...d\varphi _{v}$. \ Writing out the
dependence on $\varphi _{v},\varphi _{s}$ and  $\theta _{s}$ explicitly, the
integrals can be carried out, and the result substituted into the expression
for the current density, which in turn give the conductivity tensor. Apart
from the extra term $\propto \mathbf{v}\times \tilde{\mathbf{B}}_{1}$, and
the need to avoid some other slight simplifications for an isotropic
background, the calculations are similar to those given in Ref. \cite{Lundin2010}. Therefore we proceed directly to the final result for the
conductivity tensor, which may be written as 

\bigskip\ 

\begin{equation}
\sigma _{ij}=\sum_{\nu =+,-}\sum_{a=-\infty }^{\infty }\int \left[ \frac{%
X_{(\nu )ij}^{(\text{sp})}}{\omega -k_{z}v_{z}-a\omega _{ce}+\omega _{cg}}+%
\frac{Y_{(\nu )ij}^{(\text{cl})}+Y_{(\nu )ij}^{(\text{sp})}}{\omega
-k_{z}v_{z}-a\omega _{ce}}+\frac{Z_{(\nu )ij}^{(\text{sp})}}{\omega
-k_{z}v_{z}-a\omega _{ce}-\omega _{cg}}\right] d^{3}v  \label{sigma}
\end{equation}%
where 
\begin{equation}
Y_{(\nu )ij}^{(\text{cl})}=\frac{q_{e}^{2}}{m_{e}}\left( 
\begin{array}{ccc}
-ia^{2}\frac{\omega _{ce}^{2}\xi _{\bot }}{k_{\bot }^{2}v_{\bot }}\mathcal{J}%
_{a}^{2} & a\frac{\omega _{ce}\xi _{\bot }}{k_{\bot }}\mathcal{J}_{a}%
\mathcal{J}_{a}^{\prime } & -ia\frac{\omega _{ce}\xi _{z}}{k_{\bot }}%
\mathcal{J}_{a}^{2} \\ 
&  &  \\ 
-a\frac{\omega _{ce}\xi _{\bot }}{k_{\bot }}\mathcal{J}_{a}\mathcal{J}%
_{a}^{\prime } & -iv_{\bot }\xi _{\bot }\mathcal{J}_{a}^{^{\prime }2} & 
-v_{\bot }\xi _{z}\mathcal{J}_{a}\mathcal{J}_{a}^{\prime } \\ 
&  &  \\ 
-ia\frac{\omega _{ce}v_{z}\xi _{\bot }}{k_{\bot }v_{\bot }}\mathcal{J}%
_{a}^{2} & v_{z}\xi _{\bot }\mathcal{J}_{a}\mathcal{J}_{a}^{\prime } & 
-iv_{z}\xi _{z}\mathcal{J}_{a}^{2}%
\end{array}%
\right)   \notag
\end{equation}%
is the classical contribution, and the spin contributions are 
\begin{equation}
Y_{(\nu )ij}^{(\text{sp})}=-\mu _{e}\frac{q_{e}}{m_{e}}\left( 
\begin{array}{ccc}
0 & -\frac{\nu a\omega _{ce}\xi _{\bot }}{v_{\bot }}\mathcal{J}_{a}^{2} & 0
\\ 
&  &  \\ 
\frac{\nu a\omega _{ce}\xi _{\bot }}{v_{\bot }}\mathcal{J}_{a}^{2} & i\frac{%
\mu _{e}}{q_{e}}\frac{k_{\bot }^{2}}{\omega }\left( \frac{a\omega _{ce}}{%
v_{\bot }}\frac{\partial F_{0\nu }}{\partial v_{\bot }}+k_{z}\frac{\partial
F_{0\nu }}{\partial v_{z}}\right) \mathcal{J}_{a}^{2} & \nu k_{\bot }\xi _{z}%
\mathcal{J}_{a}^{2} \\ 
& +i\nu \left( \frac{\omega \xi _{\bot }}{v_{\bot }}+\frac{a\omega _{ce}}{%
v_{\bot }}\frac{\partial F_{0\nu }}{\partial v_{\bot }}+k_{z}\frac{\partial
F_{0\nu }}{\partial v_{z}}\right) \frac{k_{\bot }v_{\bot }}{\omega }\mathcal{%
J}_{a}\mathcal{J}_{a}^{\prime } &  \\ 
&  &  \\ 
0 & -\nu k_{\bot }\xi _{z}\mathcal{J}_{a}^{2} & 0%
\end{array}%
\right)   \notag
\end{equation}%
together with 
\begin{eqnarray}
X_{(\nu )ij}^{(\text{sp})} &=&i\frac{\mu _{e}^{2}}{\hbar \omega }\left( \nu
F_{0\nu }+\frac{\hbar }{2m_{e}}\left( \frac{a\omega _{ce}}{v_{\bot }}\frac{%
\partial F_{0\nu }}{\partial v_{\bot }}+k_{z}\frac{\partial F_{0\nu }}{%
\partial v_{z}}\right) \right) \mathcal{J}_{a}^{2}M_{ij}  \notag \\
Z_{(\nu )ij}^{(\text{sp})} &=&i\frac{\mu _{e}^{2}}{\hbar \omega }\left( -\nu
F_{0\nu }+\frac{\hbar }{2m_{e}}\left( \frac{a\omega _{ce}}{v_{\bot }}\frac{%
\partial F_{0\nu }}{\partial v_{\bot }}+k_{z}\frac{\partial F_{0\nu }}{%
\partial v_{z}}\right) \right) \mathcal{J}_{a}^{2}M_{ij}^{\ast }  \notag
\end{eqnarray}%
where 
\begin{equation}
M_{ij}=\left( 
\begin{array}{ccc}
-k_{z}^{2} & ik_{z}^{2} & k_{\bot }k_{z} \\ 
&  &  \\ 
-ik_{z}^{2} & -k_{z}^{2} & ik_{\bot }k_{z} \\ 
&  &  \\ 
k_{\bot }k_{z} & -ik_{\bot }k_{z} & -k_{\bot }^{2}%
\end{array}%
\right) .  \notag
\end{equation}%
It can be noted that the term $Y_{(\nu )ij}^{(\text{sp})}$ comes from the
spin (magnetic dipole force) contribution to the free current density,
whereas the terms $X_{(\nu )ij}^{(\text{sp})}$ and $Z_{(\nu )ij}^{(\text{sp}%
)}$ comes from the magnetization current. Since the conductivity tensor (\ref{sigma}) contain all plasma currents, the general dispersion relation is
obtained in the same way as in the classical case, i.e.\ the dispersion relation is given by $%
\text{det}D_{ij}=0$, with $D_{ij}=\delta _{ij}(1-k^{2}c^{2}/\omega
^{2})+k_{i}k_{j}c^{2}/\omega ^{2}-i\sigma _{ij}/\varepsilon _{0}\omega $.
Picking the special case of an isotropic distribution, i.e. letting $F_{0\nu
}=$ $F_{0\nu }(v^{2})$, it is straightforward to show that the conductivity
tensor reduces to the expression derived in Ref. \cite{Lundin2010}. 

Evaluating the dispersion relation is a complicated  task in its own right. Generally, the spin contribution tend to be more significant if the background magnetic fields is strong, if the plasma density is high, and if the temperature is modest. There are numerous dimensionless parameters describing this, as discussed more thoroughly in Ref. \cite{Lundin2010}. Here we would just like to point out a few more consequences of the conductivity tensor. Since there are certain denominators in the expression (\ref{sigma}) proportional to $\omega
-k_{z}v_{z}-\pm(\omega _{ce}-\omega _{cg})$, for almost perpendicular propagation (negligible $k_z$), certain spin terms will be magnified if the wave frequency matches the difference in the gyration frequency and the spin precession frequency, $\Delta \omega_c=(g/2-1)\omega_c\approx 0.00116 \omega_c$. Spin induced wave modes with  $\omega \approx \Delta \omega_c$ have been studied by e.g. Refs. \cite{Brodin2008,Asenjo-2012}. A closely related feature due to these denominators, is the spin-induced wave-particle interaction. Even if the plasma parameters are more or less classical, such that the spin terms are small, for classical wave modes with $\omega\sim \Delta \omega_c$, the smallness of the overall coefficients for the spin terms, can be compensated by a larger number of resonant particles. Spin induced damping due to this mechanism have been studied e.g. by Refs. \cite{Zamanian 2010,Ekman-2021}. For an extended treatment of the electrostatic limit, see Ref. \cite{Hussain2014}.   

Moreover, we note that since the expression for the conductivity tensor given here allows for anisotropic distributions, it may be used for studying instabilities of the Weibel type. In a quantum plasma context, such instabilities have been previously been studied by e.g. Refs. \cite{Haas2008,Weibel2018}, but without accounting for the spin dynamics. Finally, in addition to the classical free energy sources, the theory presented here allows for instabilities driven by a difference in the spin temperature and the kinetic temperature. Fig. 3, reprinted with permission from J. Lundin and G. Brodin, Phys. Rev. E {\textbf 82}, 056407 (2010), 
copyright  2010  by  the
American Physical Society, shows that a rather small difference between the spin temperature $T_{{\rm sp}}$ and the kinetic temperature $T_{\rm kin}$ is enough to drive an instability in the absence of dissipation. 
\begin{figure}\label{fig:instability}
\includegraphics[width=0.5\textwidth]{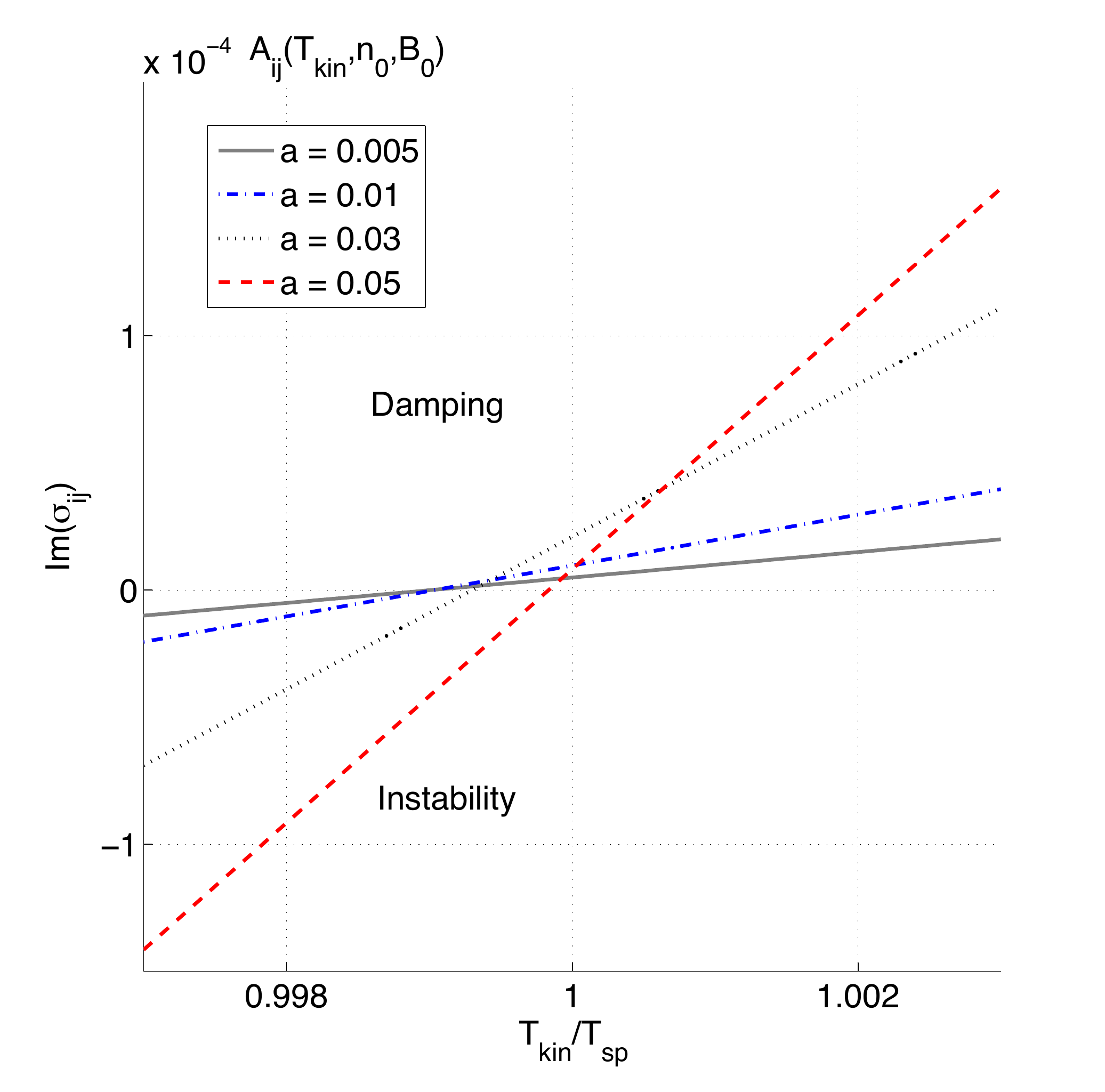}
\caption{(Color online) The imaginary contribution of $X_{ij}^{(sp)}$ to $\sigma_{ij}$ plotted as a function of $T_{\mathrm{sp}}/T_{\mathrm{kin}}$ for different values of $\tilde a$. A positive value corresponds to damping while a negative value gives rise to a instability. We refer to Ref. \cite{Lundin2010} for the detailed definitions of $\tilde a$ and $A_{ij}$ determining the instability growth rate.}
\end{figure}

\section{Conservation laws, Landau quantization, and Nonlinear effects, }
\label{section5}

Up to now, we have aimed for a clear logical structure of the review article, building up quantum kinetic theory by starting from elementary models, gradually progressing to more advanced ones. In this section, we will leave this route to some extent, and instead offer a {\it smorgasbord} of different results, in order to illustrate the diversity of quantum kinetic theory. In particular, we will consider quantum kinetic conservation laws, Landau quantization in a strong magnetic field, and we will use nonlinear perturbation theory to find the spin contribution to the ponderomotive force. Doing so, in order to present findings of a more general nature, we will apply models that to some degree extends those presented earlier. However, all of them can be derived similarly to the schemes presented above, i.e. by finding a proper Hamiltonian, using the von Neumann equation for the density matrix, and finally making a Wigner transform (and possibly also a Q-transform). The only additional feature is the need for a Foldy-Wouthuysen transform \cite{FW-transform,Silenko2008}, to isolate the electron degrees of freedom from the positron degrees of freedom. 
\subsection{Conservation laws}
Here we will use a relativistic quantum kinetic model that comes from separating positive and negative energy
solutions of the Dirac equation by means of a Foldy-Wouthuysen (F-W)
transformation~\cite{FW-transform,Silenko2008}. Since we are decoupling
electrons and positrons, the physical condition of applicability is that
pair production is negligible. Quantitatively, the electric field should be
much smaller than the critical field, $E\ll E_{c}=m^{2}/e\hbar $ and similarly
for the magnetic field. Moreover, the typical scale lengths should be long compared to the
Compton wavelength $\hbar /mc$. Apart from the F-W transformation, the derivation of the model is
similar to section III and we refer to Ref.  \cite{strongfield1} for the details.  

The evolution equation for the scalar Wigner function $f$ is 
\begin{widetext} 
	\begin{align}
	0 &  = \partial_t f + 
	\left( 
		\frac{\bf p}{\epsilon} - \mu_B \nabla_p \tilde{T}
	\right) 
	\cdot \nabla_x f 	
	+ q \left(
		{\bf E}
		+ \left( \frac{\bf p}{\epsilon} - \mu_B  \nabla_p \tilde{T} \right)
		\times \mathbf B
	\right)
	\cdot \nabla_p f
	+ \mu_B (\nabla_x\tilde{T} ) \cdot \nabla_p f
	+ \frac{2 \mu_B m }{\hbar \epsilon}  \left(
		{\bf s} \times \mathbf T
	\right)
	\cdot \nabla_s f
	\label{eq:evolution}
	\end{align} 
\end{widetext}where $\epsilon ^{2}=\mathbf{p}^{2}+m^{2}$, $\mu _{B}=q\hbar
/2m$ is the Bohr magneton and 
\begin{align}
\mathbf{T}& =\frac{m}{\epsilon }\left( \mathbf{B}-\frac{\mathbf{p}\times 
\mathbf{E}}{\epsilon +m}\right)  \\
\tilde{T}& =\mathbf{T}\cdot (\mathbf{s}+\nabla _{s}).
\end{align}

The system is closed with Maxwell's equations, in units where $c=\epsilon
_{0}=\mu _{0}=1$, 
\begin{subequations}
\begin{align}
\nabla \cdot \mathbf{E}& =\rho _{f}-\nabla \cdot \mathbf{P} \\
\nabla \cdot \mathbf{B}& =0 \\
\nabla \times \mathbf{E}& =-\frac{\partial \mathbf{B}}{\partial t} \\
\nabla \times \mathbf{B}& =\mathbf{j}_{f}+\frac{\partial \mathbf{E}}{%
\partial t}+\frac{\partial \mathbf{P}}{\partial t}+\nabla \times \mathbf{M}
\label{eq:ampere}
\end{align}%
where $\mathbf{P}$ and $\mathbf{M}$ are the polarization and magnetization
densities, and $\rho _{f}$ and $\mathbf{j}_{f}$ are the free charge and
current densities. The source terms are slightly generalized, as compared to previous models, and given by 
\end{subequations}
\begin{align}
\rho _{f}& =q\int d\Omega \,f \\
\mathbf{j}_{f}& =q\int d\Omega \,\left( \frac{\mathbf{p}}{\epsilon }-\mu
_{B}\nabla _{p}\mathbf{T}\cdot 3\mathbf{s}\right) f  \label{eq:j-free} \\
\mathbf{P}& =-3\mu _{B}\int d\Omega \,\frac{m\mathbf{s}\times \mathbf{p}}{%
\epsilon (\epsilon +m)}f  \label{eq:polarization} \\
\mathbf{M}& =3\mu _{B}\int d\Omega \,\frac{m}{\epsilon }\mathbf{s}f
\label{eq:magnetization}
\end{align}%
As the theory is relativistic,
the integration element $d\Omega $ of section III is replaced by its
relativistic counterpart, i.e. $d\Omega \,=d^{3}pd^{2}s$. It follows from
the evolution equation, Eq. (\ref{eq:evolution}), that the free charge is
conserved, $\partial _{t}\rho _{f}+\nabla \cdot \mathbf{j}_{f}=0$, and we
interpret 
\begin{equation}
\mathbf{v}=\frac{\mathbf{p}}{\epsilon }-\mu _{B}\nabla _{p}\mathbf{T}\cdot 3%
\mathbf{s}  \label{eq:velocity}
\end{equation}%
as the function on phase space corresponding to the velocity -- it is in
fact the Weyl transform of the velocity operator $\hat{\mathbf{v}}=\frac{i}{%
\hbar }[\hat{H},\hat{\mathbf{x}}]$ given by the Heisenberg equation of
motion. The spin-dependent term is related to the hidden momentum \cite
{HM1,HM2,HM3,HM4} of
systems with magnetic moments. Here we also note an important aspect of the
relativistic theory; that the spin magnetization current is complemented by
polarization currents. This is natural, of course, as a magnetic dipole
moment in the rest frame of a particle correspond to both a magnetic and
electric dipole moment in any other frame.  

The total energy density is given by 
\begin{equation}
W=\frac{1}{2}(E^{2}+B^{2})+\int d\Omega \,\left( \epsilon -3\mu _{B}m\frac{%
\mathbf{B}}{\epsilon }\cdot \mathbf{s}\right) f
\end{equation}%
with the corresponding energy flux vector 
\begin{equation}
\mathbf{K}=\int d\Omega \,\left[ \epsilon +\mu _{B}m3\mathbf{s}\cdot \left( 
\frac{\mathbf{B}}{\epsilon }-\frac{\mathbf{p}\times \mathbf{E}}{\epsilon
(\epsilon +m)}\right) \right] \mathbf{v}f \\
+\mathbf{E}\times \mathbf{H},  \label{eq:energy-flux}
\end{equation}%
where $\mathbf{H}=\mathbf{B}-\mathbf{M}$. With these expression, we have a conservation law on
divergence form 
\begin{equation}
\partial _{t}W+\nabla \cdot \mathbf{K}=0.  \label{eq:conservation-energy}
\end{equation}%
It is straightforward to confirm the above energy conservation law by
carrying out a number of partial integrations. 

Deriving the conservation law for momentum is somewhat more tedious, and we
refer to Ref. \cite{strongfield2} for the details. Here we just present the result, which
can be written in standard form in terms of energy-momentum tensors for
particles and fields: 
\begin{equation}
\partial _{t}\left( \langle \mathbf{p}\rangle +\mathbf{D}\times \mathbf{B}%
\right) _{i}+\partial _{x_{j}}(T_{ij}^{\text{e}}+T_{ij}^{\text{EM}})=0.
\end{equation}%
Here the electromagnetic part of the energy-momentum tensor $T_{ij}^{\text{EM%
}}$ is given by%
\begin{equation}
T_{ij}^{\text{EM}}=\frac{1}{2}\left( E^{2}+B^{2}-2\mathbf{M}\cdot \mathbf{B}%
\right) \delta _{ij}-H_{i}B_{j}-E_{i}D_{j}.
\end{equation}%
where the usual relations $\mathbf{D=E+P}$ and $\mathbf{B}=\mathbf{H}-%
\mathbf{M}$ applies. Similarly, the electron contribution $T_{ij}^{\text{e}}$
to the energy-momentum tensor is given by 
\begin{equation}
T_{ij}^{\text{e}}=\int d\Omega \,p_{i}v_{j}f.  \label{particle}
\end{equation}%
where  the momentum-velocity relation (\ref{eq:velocity}) apply. 

A few things can be noted. Firstly, the corresponding conservation laws for
the non-relativistic model in section III can be obtained as expected from the above results, i.e. by dropping terms of higher order in an expansion ${\bf p}/m$ (letting $\epsilon = m(1+p^2/2)$, etc.) Secondly, with an expression for the energy momentum tensor, in
principal we can compute the gravitational source due to quantum
relativistic electrons. In practice, this is complicated by the fact that
the stress tensor is not necessarily symmetric, see Ref. \cite{strongfield2} for a more
thorough discussion. Finally, we note that the conserved quantities will not be
modified due to our neglect of short scale effects (of the type contained in Eqs. (\ref{Etilde})-(\ref{dBtilde})). The reason is that all the extra terms due to short scale effects
contain higher order derivatives whose contributions to the energy momentum
tensors vanish when integrating over momentum space.  
\subsection{Landau quantization}
When the Zeeman energy in an external magnetic field is large, i.e. comparable to the kinetic energy of particles, the phenomenon of Landau quantization becomes crucial. This means that the energy levels for motion perpendicular to a magnetic field are quantized, and also that the energy difference between the spin states are significant. In particular this tend to occur in astrophysical scenarios, where, in extreme cases, the Zeeman energy may be comparable to the electron rest mass energy. Specifically this happens in the vicinity of magnetars, where the magnetic field strength can be of the order $10^{10}$ ${\rm T}$ \cite{magnetar}. Here we will address the regime of fully relativistic Landau quantization, when the effect is most pronounced. A particular difficulty with relativistic Landau quantization, is that we can no longer use the inequality $\mu_B B \ll mc^2$, (or similarly for the electric field) which plays an important role when separating electron and positron states. In principal, this could be handled using the DHW-formalism to be discussed in the next section, but this comes at the price of a considerably more complex theory. In order to focus solely on the problem of Landau quantization, we here takes a simpler route, and consider a strong constant magnetic field $B_0$ (allowing for $\mu_B B_0 \sim mc^2$ or even larger), but limit the magnitude of electromagnetic field perturbations well beyond this.    

A model focusing on the effect of Landau quantization may still neglect
spin-dynamics, in case the validity condition $\hbar k^{2}/m\omega \ll 1$ is
fulfilled. (Here $k$ and $\omega $ do not necessarily refer to plane waves,
instead they represent characteristic spatial and temporal gradients.) Also
assuming $\hbar k\nabla _{p}\ll 1$, the previously studied short scale
effects can be dropped, and the model will be a slight generalization of the
relativistic Vlasov equation, extended to account for Landau quantization in
a strong (constant) magnetic field $B_{0}$. The kinetic evolution equation
derived in Ref. \cite{landau-quant} can be written 
\begin{equation}
\partial _{t}W_{\pm }+\frac{1}{\epsilon _{\pm }^{\prime }}\mathbf{p}\cdot
\nabla _{r}W_{\pm }+q\Big[\mathbf{E}+\frac{1}{\epsilon _{\pm }^{\prime }}%
\mathbf{p}\times \mathbf{B}\Big]\cdot \nabla _{p}W_{\pm }=0.
\label{Land-quant}
\end{equation}%
where the main difference to the (relativistic) Vlasov equation lies in the
energy expression $\epsilon _{\pm }^{\prime }$, which now becomes an
operator given by 
\begin{equation}
\epsilon _{\pm }^{\prime }=\sqrt{m^{2}+\mathbf{p}^{2}\mp 2m\mu
_{B}B_{0}-m^{2}\mu _{B}^{2}(\mathbf{B}_{0}\times \nabla _{p})^{2}}.
\label{energy_operator}
\end{equation}%
Here the first two terms inside the root sign just give the classical
expression. The next term with a $\mp $ sign gives the magnetic dipole
energy for spin up and spin down respectively.Accordingly, the upper and
lower signs are described by different evolution equations for $W_{+}$ and $%
W_{-}$, due to the different energies of the spin up and spin down
particles. The final non-classical feature comes from the last term under
the root sign, which makes the energy an operator instead of just an
algebraic expression.  Finally, to have a closed system, we need the source
terms in Maxwell's equations, which are given by summing over the spin up
and down contribution, i.e. 
\begin{align}
\rho _{f}& =q\sum_{\pm }\int d^{3}p\,W_{\pm } \\
\mathbf{j}_{f}& =q\sum_{\pm }\int d^{3}p\,\frac{1}{\epsilon _{\pm }^{\prime }%
}\,\mathbf{p}W_{\pm }.
\end{align}%
In principle, there are also magnetization currents that can be added to the
free sources above. However, such a contribution will be negligible in
comparison, provided the condition given above for neglecting the magnetic
dipole force is fulfilled. 

Next we want to deduce the thermodynamic equilibrium state in a constant
magnetic field. Noting that for a constant magnetic field, both the Dirac
equation and the Pauli equation results in electrons obeying a quantum
harmonic oscillator equation, we can make a straightforward generalization
of the Pauli case \cite{Zamanian 2010}. Both for the Pauli and the Dirac
equations, the spatial dependence of the wavefunction in Cartesian
coordinates can be expressed as a Hermite polynomial times a Gaussian
function \cite{melrose1983} only the energy eigenvalues for the Landau
quantized states are different. Specifically, applying the Dirac theory, the
energy of the Landau quantized states become%
\begin{equation}
E_{n\pm }=m\sqrt{1+(2n+1\pm 1)\frac{\hbar \omega _{ce}}{m}+\frac{p_{z}^{2}}{%
m^{2}}}  \label{relativisticenergy}
\end{equation}%
where $n=0,1,2,\ldots $ corresponds to the different Landau levels for the
perpendicular contribution to the kinetic energy, the index $\pm $
represents the contribution from the two spin states, and the term
proportional to $p_{z}^{2}$ gives the continuous dependence on the parallel
kinetic energy. Since the Pauli and Dirac equations for individual particle
states have the same spatial dependence for the wave function, we can adopt
the expression for the Wigner function from Ref.~\cite{Zamanian 2010} (based
on the Pauli equation) with some relatively minor adjustments.

\begin{enumerate}
\item Contrary to Ref.~\cite{Zamanian 2010}, we have made no Q-transform to
introduce an independent spin variable, and thus the spin-dependence of Ref.~%
\cite{Zamanian 2010} reduces to $W_\pm$.

\item The Wigner function of Ref.~\cite{Zamanian 2010} must be expressed in
terms of the momentum, i.e., $m(v_{x}^{2}+v_{y}^{2})/2 \rightarrow
(p_{x}^{2}+p_{y}^{2})/2m$. 

\item The non-relativistic energy of Ref.~\cite{Zamanian 2010} is replaced by
the relativistic expression (\ref{relativisticenergy}) of the Dirac theory. 

\item The normalization of the Wigner function must be adopted to fit the
present case.
\end{enumerate}

With these changes, the background Wigner function $W_{\pm}^{TB}$ for the
case of electrons in thermodynamic equilibrium can be written%
\begin{equation}
W_{\pm}^{TB} = \frac{n_{0 \pm}}{\left( 2 \pi \hbar \right)^3} \sum_{n} \frac{%
2(-1)^n \phi_n (p_{\perp})}{\exp \left[ (E_{n,\pm }-\mu_{c})/k_{B}T\right] +
1},  \label{Wigner-thermo}
\end{equation}
where 
\begin{equation}
\phi_n(p_\perp) = \exp \left( -\frac{p_{\perp}^2}{m\hbar \omega_{ce}^2}
\right) L_n\left( \frac{2p_{\perp}^2}{m\hbar \omega_{ce}^2}\right),
\label{Eigen-func}
\end{equation}
$n_0=n_{0+}+n_{0-} =\int (W_{+T}+W_{-T}) d^3 p$ is the electron number
density of the plasma, $\mu_{c}$ is the chemical potential, $T$ is the
temperature, and $L_n$ denotes the Laguerre polynomial of order $n$.

That the factor $\phi_n(p_{\perp}) $ gives us the proper Wigner function for
the Landau quantized eigenstates can be confirmed by an independent check.
Since the expression (\ref{Wigner-thermo}) contains no dependence on the
azimuthal angle in momentum space, we can write 
\begin{widetext}
    \begin{equation}
        \epsilon^{\prime}_{\pm}
        =
        m\sqrt{
            1 + p_\perp^2/m^2
            - \mu_B^2 B_{0}^{2}\left(
                \frac{\partial }{\partial p_\perp} + \frac{1}{p_\perp}
            \right) \frac{\partial }{\partial p_\perp}
            \mp \frac{2\mu_B B_0}{m} + \frac{p_z^2}{m^2}
        }
        \label{simplified expression}
    \end{equation}
\end{widetext}
when $\epsilon^{\prime }_{\pm}$ acts on $\phi_n(p_\perp)$. Computing $%
\epsilon^{\prime }_{\pm}\phi_n(p_\perp)$ by Taylor-expanding the square root
to infinite order, using the properties of the Laguerre polynomials, and
then converting the sum back to a square-root, it is straightforward to
verify the relation 
\begin{widetext}
    \begin{equation}    
        \label{energy_operator2}
        \epsilon^{\prime}_{\pm} \phi_n(p_\perp) 
        =
        m\left(
            1 + (2n + 1 \pm 1) \frac{\hbar \omega_{ce}}{m}+\frac{p_{z}^{2}}{m^{2}}
        \right)^{1/2}
        \phi_n(p_\perp) 
    \end{equation}
\end{widetext}
where $\omega_{ce} = \frac{|qB_0|}{m}$ is the electron cyclotron frequency,
confirming that $\phi_n(p_\perp) $ generates the proper energy eigenvalues
for the perpendicular kinetic energy and the spin degrees of freedom.

While (\ref{Wigner-thermo}) gives the thermodynamic equilibrium expression 
$W_{\pm }^{TB}$, we note that the plasma background state is not necessarily
in thermodynamic equilibrium. Making use of the property 
(\ref{energy_operator2}), 
we note that the most general time-independent
solution $W_{0\pm }$ to (\ref{Land-quant}) of physical significance
can be written in the form 
\begin{equation}
W_{0\pm }=\sum_{n}g_{n,\pm }(p_{z})(-1)^{n}\phi _{n}(p_{\perp })
\label{Background-general}
\end{equation}
where $g_{n\pm }(p_{z})$ is a function that is normalizable, but otherwise
arbitrary, and the number of particles in each Landau quantized eigenstate 
$n_{n,\pm }$ obeys the condition 
\begin{align}
n_{n\pm }=& \int g_{n\pm }(p_{z})(-1)^{n}\phi _{n}(p_{\perp })\,d^{3}p 
\notag \\
\Rightarrow &   \notag \\
n_{n\pm }=& \frac{(2\pi \hbar )^{3}}{2}\int g_{n\pm }(p_{z})dp_{z}.
\label{Norm-condition}
\end{align}
Naturally, the expressions for $W_{0\pm }$ and $W_{\pm }^{TB}$ presented
here are of most significance for relativistically strong magnetic fields,
when Landau quantization is pronounced. As a consequence, the above formulas
will reduce to more well-known expressions when the limit $\hbar \omega
_{ce}/m\ll 1$ is taken. Specifically, Eq. (\ref{Wigner-thermo}) will become a
relativistically degenerate Fermi-Dirac distribution in case we let $T=0$
and $\mu _{c}=E_{F}\gg \hbar \omega _{ce}$, where $E_{F}$ is the Fermi
energy. Alternatively, for $k_{B}T\gg E_{F}$ and $k_{B}T\gg \hbar \omega_{ce}$, Eq. (\ref{Wigner-thermo}) reduces to a Synge-Juttner distribution. 

\begin{figure}
    \centering
    \includegraphics[width=0.7\textwidth]{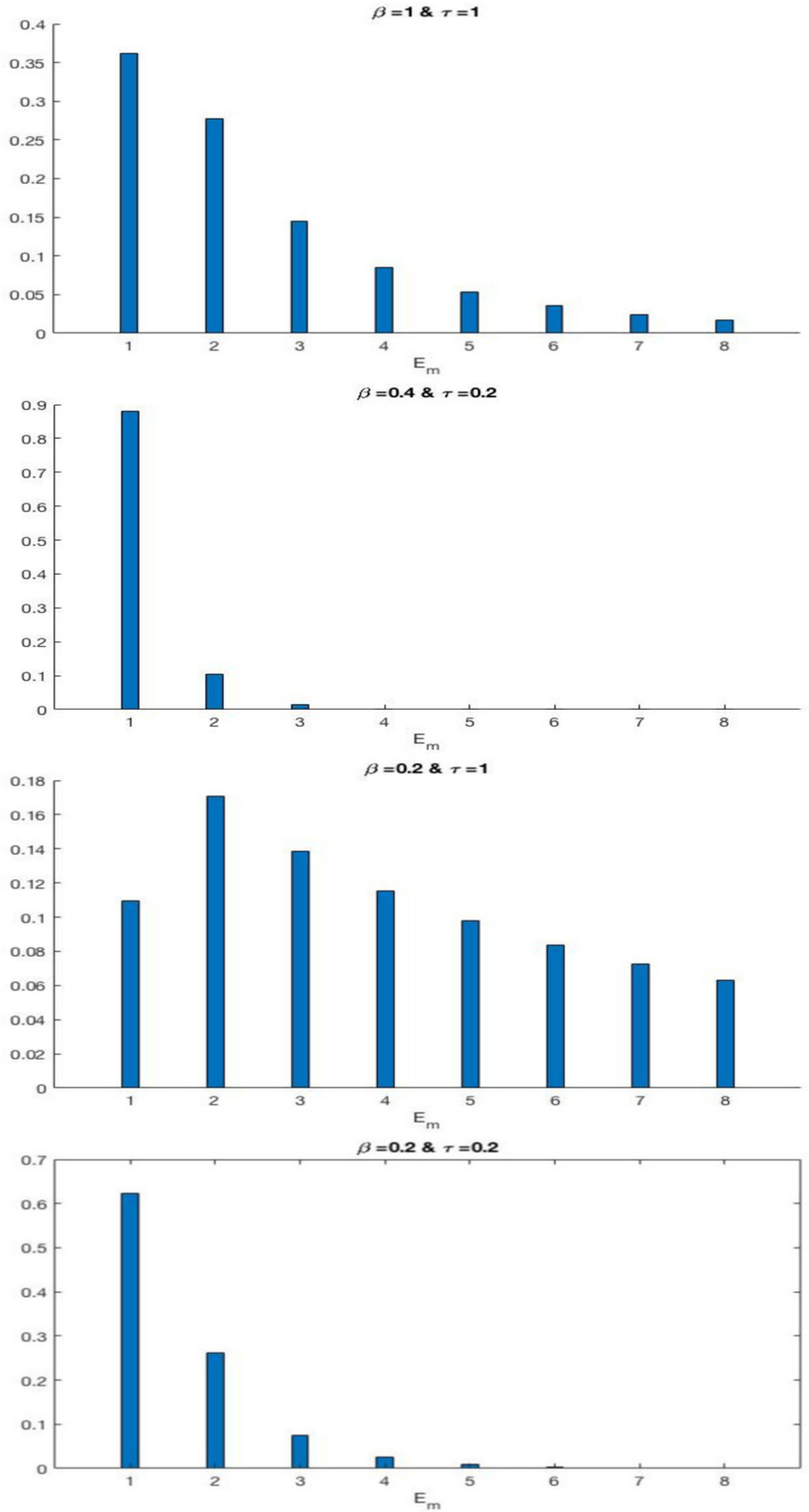}
    \caption{The normalized number density at different energy states $E_m$ for different values of the parameters $\beta=\mu_BB_0/m$ and $\tau= k_{BT}/m$.}
	\label{fig:background}
\end{figure}

To give a concrete illustration, in Fig. IV, reprinted with permission from \cite{Al-Naseri-2021} copyright  2021  by  the American Physical Society, we show
a bar chart for the normalized number density $n_{0n\pm}/n_0$
in the different energy states, for a few values of the temperature and magnetic field, under the assumption that the density is low enough for the system to be non-degenerate, i.e. assuming $T>T_F$.

As a result of the background dependence, Eq. (\ref{Background-general}), in the
Landau quantized regime, the electrons behave as a multi-species system,
where each particle species has its own rest mass, as given by 
Eq. (\ref{relativisticenergy}) but with $p_{z}=0$. This is because the separation
between Landau levels is of the order of the rest mass, and all excitations by
quanta with energies of that order have been neglected. If we define the
effective number density of each "species" (discrete energy-state) as 
\begin{equation}
n_{0n\pm }\equiv \frac{n_{0}}{\left( 2\pi \hbar \right) ^{3}}\int d^{3}p%
\frac{2\left( -1\right) ^{n}\phi _{n}(p_{\perp })}{\exp \left[ \left(
E_{n\pm }-\mu _{c}\right) /k_{B}T\right] +1},
\end{equation}%
we see that $n_{0n\pm }$ to a large degree will be determined by the
Boltzmann factors of Eq. (\ref{Wigner-thermo}). For a study of Langmuir waves in
a Landau quantized plasma, see Ref. \cite{landau-quant}. 

While we have here focused on the extreme case of relativistic Landau quantization, we note that the thermodynamics is much affected also in the non-relativistic regime, provided the Zeeman energy is of the same magnitude or larger than the characteristic kinetic energy in the background plasma. For applications of Landau quantization in the non-relativistic regime, see e.g. \cite{LQ1,LQ2}

\subsection{Nonlinear effects}
Not surprisingly, quantum kinetic models can describe a great variety of nonlinear phenomena. Just like for classical plasmas, a fair share of these phenomena is induced by the ponderomotive force. For example, the ponderomotive force is the driver of plasma wake field generation \cite{Wakefield}, the key mechanism in soliton formation \cite{Soliton}, and the main source of nonlinear self-focusing \cite{Selffocusing}. While the main features of classical and quantum kinetic models driven by the ponderomotive force are similar, nevertheless there are important differences. For one thing, in a magnetized plasma, the classical ponderomotive force have  cyclotron resonances \cite{Karpman}. In quantum kinetic theory, the classical terms are still present, but they are complemented by terms containing extra quantum resonances \cite{PRL2010,PRE2011}. 

The concept of a ponderomotive force in quantum kinetic theory is not as straightforward as in a fluid theory. Depending on definition, some low-frequency nonlinearities induced by quadratic nonlinearities may be included as a ponderomotive force term or not. In order to avoid ambiguities, we will focus on the regime where the phase velocity of the high-frequency wave is much higher than the thermal velocity (or characteristic velocity, in case of degeneracy effects). 

As a starting point, we will base our study on the long-scale version of the
model put forward in Section  III, that is, Eq. (\ref{paulilong}). Next,
we consider circularly polarized electromagnetic waves of high frequency
propagating parallel to an external magnetic field, $\mathbf{B}_{0}=B_{0}%
\hat{z}$, and use the following ansatz 
\begin{align}
\mathbf{E}& =\frac{1}{2}\left[ \tilde{\mathbf{E}}(z,t)e^{i\left( kz-\omega
t\right) }+\tilde{\mathbf{E}}^{\ast }(z,t)e^{-i\left( kz-\omega t\right) }%
\right] ,  \notag \\
\mathbf{B}& =\frac{1}{2}\left[ \tilde{\mathbf{B}}(z,t)e^{i\left( kz-\omega
t\right) }+\tilde{\mathbf{B}}^{\ast }(z,t)e^{-i\left( kz-\omega t\right) }%
\right] .  \label{Ansatz-1}
\end{align}%
The amplitudes are assumed to vary much slower than the exponential phase
factors, and the star denotes complex conjugates. Since the basic wave modes
propagating parallel to $\mathbf{B}_{0}$ are either left- or
right-circularly polarized, we have $\tilde{\mathbf{E}},\tilde{\mathbf{B}}%
\propto $ $\widehat{\mathbf{x}}\pm i\widehat{\mathbf{y}}$. Furthermore, all
perturbations are small, such that weakly nonlinear perturbation theory is
applicable, and we will focus on the ponderomotive contribution, that is the
quadratically nonlinear low-frequency terms.  

In order to calculate the weakly nonlinear low-frequency response to an
incoming transverse wave packet we make the ansatz 
\begin{align}
f(\mathbf{x},\mathbf{v},\mathbf{s},t)=& f_{0}(v^{2},\theta _{s})+f_{\mathrm{%
lf}}(z,t,\mathbf{v,}\theta _{s})  \notag \\
& +\frac{1}{2}\left[ f_{1}(z,t,\mathbf{v},\mathbf{s})e^{ikz-i\omega
t}+f_{1}^{\ast }(z,t,\mathbf{v},\mathbf{s})e^{-ikz+i\omega t}\right] ,
\label{Dist_ansatz}
\end{align}%
where $f_{0}$ is the background distribution, $f_{\mathrm{lf}}$ is a
low-frequency part due to quadratic nonlinearities and $\tilde{f}_{1}$ is a
slowly modulated high-frequency wave. The background distribution will be
taken to be of the form 
\begin{equation}
f_{0}=\frac{n_{0}}{2\pi ^{3/2}v_{T}^{3}}e^{-v^{2}/v_{T}^{2}}\left[ 1+\tanh
\left( \frac{\mu B_{0}}{k_{B}T}\right) \cos \theta _{s}\right] \label{Background},
\end{equation}%
where $n_{0}$ is the equilibrium density, the thermal velocity $v_{T}$ is
defined as $v_{T}=\sqrt{2k_{B}T/m}$. It should be stressed that the
background distribution (\ref{Background}) has been picked mostly for convenience. In
fact, the results below are not depending sensitively on the background, as
long as the phase velocity is larger than the characteristic velocity. For
example, a degenerate distribution, but with the Fermi velocity smaller than
the phase velocity would produce identical results.  

Given the ansatz \eqref{Ansatz-1} and (\ref{Dist_ansatz}), the aim is then
to find an equation for the low-frequency part of the distribution function.
From such an equation we can then calculate the low-frequency response in
the current density and magnetization, and compare with the results for a
ponderomotive force, as defined from a fluid theory. First we deduce the
high-frequency linear perturbation of the distribution function, which is
given by $f_{1}=f_{+}$ ($f_{-})$ for left-hand (right-hand) circularly
polarized waves. The expressions for $f_{\pm }$ found from Eq.\ \eqref{paulilong}
to linear order can be easily computed from Eq. (9) in Ref.\ \cite{Lundin2010}, and the result is 
\begin{equation}
f_{\pm } = \frac{(-i)e^{\mp i\varphi _{v}}}{\omega -kv_{z}\mp \omega _{c}}%
\frac{q}{2m}E_{\pm }\partial _{v_{\perp }}f_{0}  
+\frac{e^{\mp i\varphi _{s}}}{\omega -kv_{z}\mp \omega _{cg}}\frac{\mu }{2m%
} \times \left[ kB_{\pm }\left( \sin \theta _{s}\partial _{v_{z}}f_{0}+\cos
\theta _{s}\partial _{\theta _{s}}\partial _{v_{z}}f_{0}\right) \pm \frac{2m%
}{\hbar }B_{\pm }\partial _{\theta _{s}}f_{0}\right] .   \label{HFlinear}
\end{equation}%
Next, allowing for slow modulations and solving the equation to first order
in $\partial _{z}/k$, $\partial _{t}/\omega $, we note that the zero-order
solution applies after making the substitution $\omega \rightarrow \omega
+i\partial _{t}$ and $k\rightarrow k-i\partial _{z}$ in Eq.\ \eqref{HFlinear},
and then expanding to first order in the slow derivatives. Inserting the
ansatz above into the evolution equation and consider the slow-time scale
and keeping only up to quadratic nonlinearities we obtain the equation 
\begin{eqnarray}
&&\left( \partial _{t} + v_{z} \partial _{z}\right) f_{lf}+\frac{q}{m}%
E_{zlf}\partial_{v_{z}}f_{0}  \notag \\
&=& -\left[ \frac{q}{4m}\left( \tilde{\mathbf{E}}+\mathbf{v}\times \tilde{%
\mathbf{B}} \right) +\frac{\mu }{4m}\left( ik+\partial _{z}\right) \left( 
\mathbf{s}\cdot \tilde{\mathbf{B}}+\tilde{\mathbf{B}}\cdot \nabla
_{\hat{\mathbf{s}}} \right) \hat{\mathbf{z}}\right] \cdot \nabla_{\mathbf{v}} \tilde{f}_{1}^{\ast } 
\notag \\
&&\quad -\frac{\mu }{2\hbar }\mathbf{s}\times \tilde{\mathbf{B}}\cdot \nabla_{\hat{\mathbf{s}}} \tilde{f}_{1}^{\ast }+\mathrm{c.c.}
\label{fpm}
\end{eqnarray}%
where c.c. stands for complex conjugate. Here we have also added a low-frequency electric field in the $z$-direction, 
$E_{z\mathrm{lf}}$, which has $f_{\mathrm{lf}}$ as source. Equations (\ref{fpm}) and (\ref{HFlinear}) now constitute a basis for calculating the nonlinear
response in the current density and magnetization. 

After some algebra (described more closely in Ref. \cite{PRE2011}), based on the
low-frequency part of Ampere's law, we end up with the final expression for
the induced low-frequency field, which can be written in terms of the
ponderomotive force $f_{p}$ as%
\begin{equation}
\left[ \partial _{t}^{2}+\omega _{p}^{2}\right] E_{z\mathrm{lf}}=\frac{qn_{0}%
}{\varepsilon _{0}}f_{p}  \label{Amp-pond}
\end{equation}%
where $f_{p}$ can be divided into its classical and its spin contribution
according to%
\begin{equation*}
f_{p}=f_{p}^{\mathrm{cl}}+f_{p}^{\mathrm{sp}}
\end{equation*}%
with the different parts given by%
\begin{equation}
f_{p}^{\mathrm{cl}}=\frac{\omega _{p}^{2}\omega }{8n_{0}mk^{2}(\omega \mp
\omega _{c})}\left( \partial _{z}\mp \frac{\omega _{c}k}{\omega (\omega \mp
\omega _{c})}\partial _{t}\right) \left\vert B_{\pm }\right\vert ^{2}
\label{pond_class}
\end{equation}%
and%
\begin{align}
f_{p}^{\mathrm{sp}}=& -\frac{\omega _{p}^{2}\hbar ^{2}k^{2}}{%
16n_{0}m^{3}(\omega \mp \omega _{cg})^{2}}\left( \partial _{z}+\frac{2k}{%
\omega \mp \omega _{cg}}\partial _{t}\right) \left\vert B_{\pm }\right\vert
^{2}  \notag \\
& \mp \frac{\hbar \omega _{p}^{2}}{8n_{0}m^{2}(\omega \mp \omega _{cg})}%
\tanh \left( \frac{\mu B_{0}}{k_{B}T}\right) \left( \partial _{z}-\frac{k}{%
\omega \mp \omega _{cg}}\partial _{t}\right) \left\vert B_{\pm }\right\vert
^{2}.  \label{pond_spin2}
\end{align}
The classical ponderomotive term (\ref{pond_class}) agree with an expression
first derived by Ref. \cite{Karpman}. \ The spin ponderomotive term (\ref{pond_spin2}%
)was first derived in Ref. \cite{PRE2011}. Our spin term agree with that work,  although
- in contrast to that expression - we have integrated the contribution over
the spin-up and the spin-down term. Note that although the modest scale
lengths assumed tend to limit the magnitude of the spin terms, the quantum
contribution can still be larger or comparable to the classical terms, in
particular close to resonances  $\omega \simeq \omega _{c}$ or $\omega
\simeq \omega _{cg}$. Note that due to the closeness of the spin-precession
frequency $\omega _{cg}$ and the cyclotron frequency $\omega _{cg}$, these
resonances can be more or less overlapping. 

An interesting feature of the above expression is the second term in (\ref%
{pond_spin2}), containing the $\tanh -$factor. If we had not averaged over
the spin states, spin-up and spin-down electrons would be affected in
opposite direction (recall that we get the $\tanh -$factor from the
difference in the number of spin-up and down particles), and hence a part of
the spin contribution to the ponderomotive force leads to a
spin-polarization being induced by the high-frequency wave. We will not
explore the consequences of the expression (\ref{pond_spin2}) further,
however, but refer to Refs \cite{PRL2010,PRE2011,PRE2010,POP2010}
that have studied various types of
nonlinear dynamics induced by the spin ponderomotive force. 

Before ending the description of spin ponderomotive physics, it is worth
noting that in certain cases, weakly relativistic contributions, in
particular the spin-orbit correction, can be important for the end result.
To explore this fact, we will make use of the weakly relativistic limit of
Eq. (\ref{eq:evolution}). obtained by letting $\varepsilon \rightarrow m$. This
model was first derived by Ref. \cite{Asenjo-2012}, and is given by  
\begin{align}
0 & = \frac{\partial f}{\partial t} + 
\left\{ \frac{\mathbf{p}}{m} + \frac{\mu}{2mc} \mathbf{E} 
\times (\mathbf{s} + \nabla_{\hat{\mathbf{s}}}) \right\} \cdot \nabla_{\mathbf{x}} f + q 
\left( \frac{1}{c} \left\{ \frac{\mathbf{p}}{m} + \frac{\mu }{2mc} \mathbf{E} \times 
(\mathbf{s} + \nabla_{\hat{\mathbf{s}}}) \right\} \times \mathbf{B} + \mathbf{E} \right) \cdot \nabla_{\mathbf{p}}f  \notag \\
& +\frac{2\mu }{\hbar} \mathbf{s} \times \left( \mathbf{B} - \frac{\mathbf{p} \times \mathbf{E}}{2mc}%
\right) \cdot \nabla_{\hat{\mathbf{s}}} f + \mu \left( \mathbf{s} + \nabla_{\hat{\mathbf{s}}} \right) \cdot 
\left[\partial_{x}^{i}\left( \mathbf{B} - \frac{\mathbf{p} \times \mathbf{E}}{2mc}\right) \right]
\partial_{p}^{i}f.  
\label{1}
\end{align}%
While we still use put gamma factors to unity (or $\varepsilon \rightarrow m$%
), the model keep terms linear in a $v/c$-expansion, such that Thomas
precession, spin-orbit terms, and a non-trivial velocity-momentum relation
(including spin) is kept. Taking the appropriate limit, the source terms used
in Section III is replaced by 
\begin{align}
\rho_{T} &= \rho_{F} + \nabla_\mathbf{x} \cdot \mathbf{P}, \\
\bm J_{T} &= \mathbf{J}_{F} + \nabla_\mathbf{x} \times \mathbf{M} + \frac{\partial \mathbf{P}}{\partial t} .
\end{align}%
Here $\rho _{F}=q\int \!d\Omega \,f$ is the free charge density and the free
current density, the polarisation and magnetisation are given by 
\begin{align}
\mathbf{J}_{f} &= q \int \!d \Omega \, \left( \frac{\mathbf{p}}{m}+\frac{3\mu}{2mc}
\mathbf{E} \times \mathbf{s} \right) f, \\
\mathbf{P} & = -3\mu \int \!d\Omega \,\frac{\mathbf{s} \times \mathbf{p}}{2mc} f, \\
\mathbf{M} & =3\mu \int \!d\Omega \,\mathbf{s} f.
\end{align}%
Next, we make the same calculation as for the simpler (non-relativistic)
model (given by Eq. (\ref{paulilong})), considering a quasi-monchromatic circularly
polarized electromagnetic wave. However, to reduce the algebra we consider a
non-magnetized plasma, i.e. we let $B_{0}\rightarrow 0$. We refer the reader
to Ref.  \cite{POP2013} for the algebraical details, and move on to the end result for
the ponderomotive force.  \bigskip 
\begin{equation}
\left( \frac{\partial ^{2}}{\partial t^{2}}%
+\omega _{p}^{2}\right) E_{lf}=\frac{qn_{0}f_p}{\varepsilon _{0}}  \label{pond_3}
\end{equation}%
where, due to the technicalities of the model, we have a result for the
time-derivative of the ponderomotive force, rather than the ponderomotive
force $f_p$ itself. The expression is 
\begin{align}
\frac{\partial f_{p}}{\partial t}=& \frac{32}{3}\frac{\pi ^{2}\mu ^{2}k^{2}}{%
m^{2}\omega ^{2}}\Bigg\{\left[ \frac{11}{2}\left( 1-\frac{k^{2}c^{2}}{\omega
^{2}}\right) \frac{\partial }{\partial z}+\frac{1}{2c}\left( 1+\frac{kc}{%
\omega }-\frac{2\omega }{kc}\right) \frac{\partial }{\partial t}-\frac{%
4k^{2}c^{2}}{\omega ^{2}}\frac{\partial }{\partial z}-\frac{2k^{2}c^{2}}{%
\omega ^{2}}\left( \frac{\partial }{\partial z}+2\frac{k}{\omega }\frac{%
\partial }{\partial t}\right) \right.   \notag \\
& +\left. \frac{3}{2}\frac{\omega }{c^{2}k}\left( 1+\frac{k^{3}c^{3}}{\omega
^{3}}\right) \frac{\partial }{\partial t}\right] \frac{\partial }{\partial t}%
+3\frac{kc^{2}}{\omega }\frac{\partial ^{2}}{\partial z^{2}}\Bigg\}%
(|E_{x}|^{2}+|E_{y}|^{2}).
\end{align}%
The last term is the classical contribution, and the second to last term in
the square bracket is what is obtained without the weakly relativistic
effects. However, if we assume that $kc/\omega $ is roughly of order unity,
we see that all terms in the square brackets are of the same order. This
implies that when dealing with an unmagnetized plasma where spin effects are
important, the spin orbit coupling contributions must be taken into account
as well.  However, the previous result for the ponderomotive force, Eq.
(\ref{pond_spin2}) is still relevant, as spin terms tends to be more important in
magnetized plasmas, and hence the simpler model (\ref{paulilong}) can still be
justified.

\section{The full Dirac theory - the DHW equations}
\label{section6}

Quantum relativistic treatments are of interest in several different contexts \cite{QRP-1,QRP-2,QRP-3}. 
Dense astrophysical objects can have a Fermi energy approaching or exceeding the electron rest mass energy, the strong magnetic fields
of magnetars give rise to relativistic Landau quantization . Importantly,  the continuous evolution
of laser intensity brings a variety of quantum relativistic
phenomena accessible to experimentalists. Upcoming laser
facilities of interest in this context include, e.g., the extreme
light infrastructure (ELI) \cite{Eli, Dunne} and the European x-ray free
electron laser (XFEL) \cite{XFEL,Ringwald},

The quantum kinetic models of previous sections has all made various simplifications as compared to the full quantum relativistic theory. In particular, in order to avoid the mixed electron-positron states of the Dirac theory, up to now we have imposed limitations on the electric field. Specifically, we have demanded $E\ll E_{\rm cr}$ where $E_{\rm cr}=m^2c^3/e\hbar$ is the critical field, in order to avoid the complications associated with significant pair-production due to the Schwinger mechanism. In this section, however, we will take on the full complexity of the Dirac theory using the so called Dirac-Heisenberg-Wigner (DHW) formalism \cite{BB-1991}. Compared to previous sections, new features of the theory includes, Zitterbewegung (a rapid (speed of light) particle motion associated with interference between positive and negative energy states), vacuum polarization, and electron-positron pair creation. Also, the previous 2 by 2 Wigner matrix will be replaced by 16 components, due to the 4 components of the Dirac spinors. Nevertheless, many other aspects of the quantum kinetic theory will be familiar and we will point out how the DHW-formalism can be related to the quantum-kinetic approximations presented earlier.  

\subsection{The DHW-model}
The DHW-model was first derived by Ref. \cite{BB-1991}. Moreover, some relatively minor variations of this derivation has been published in the literature more recently, see e.g. \cite{Gies,Sheng}. As all these treatments are fully satisfactory, we will not repeat the calculations, but just point out a few of the main features.  

\begin{enumerate}
    \item The derivation is based on the Dirac equation, which gives the time evolution of Dirac four spinors, generally describing mixed electron positron states.  
    \item Just like in previous theories, a gauge-invariant Wigner transformation is made, which here produce  a 4 by 4 Wigner matrix, with the components depending on phase-space variables just like in standard kinetic theory.
    \item The main omission is made when using  the Hartree-approximation where the electromagnetic field is treated as a non-quantized field.  This approximation amounts to neglecting the quantum fluctuations. We will come back to the consequences of this approximation. 
    \item In order to write the equations in a physically more transparent form, the 16 components of the Wigner matrix is decomposed into 4 different four vectors, which in turn is split into temporal and spatial components. Most of these quantities have a clear physical meaning, which helps forming a physical understanding of the theory.
    \item The (phase-space) current density and charge density are parts of the DHW-functions, which makes it straightforward to close the system using Maxwell's equations. 
\end{enumerate} 
With these preliminaries, we jump directly to the DHW-equations, which in units where $c=\hbar=1$ can be written in the form: 
\begin{align}
    D_t s-2\Tilde{\textbf{p}}\cdot \textbf{t}_1&=0 \notag\\
    D_t\varrho+2 \Tilde{\textbf{p}}\cdot \textbf{t}_2 &=2ma_0\notag\\
    D_tv_0 +\textbf{D} \cdot \textbf{v}&=0\notag \\
    D_ta_0 + \textbf{D}\cdot \textbf{a}&=-2m \varrho \notag \\
    D_t \textbf{v} + \textbf{D}v_0 - 2\Tilde{\textbf{p}}\times \textbf{a}&=-2m\textbf{t}_1\notag \\
    D_t\textbf{a} + \textbf{D}a_0 - 2\Tilde{\textbf{p}}\times \textbf{v}&=0\notag \\
    D_t\textbf{t}_1+ \textbf{D}\times \textbf{t}_2 + 2\Tilde{\textbf{p}} s&=2m\textbf{v} \notag \\
    D_t \textbf{t}_2- \textbf{D}\times \textbf{t}_1 
    -2\Tilde{\textbf{p}}\varrho &=0. 
    \label{DHW}
\end{align}
Due to the Wigner transform, we cover short-scale quantum phenomena in much the same way as in the previous sections. This is illustrated by the appearance of the non-local operators in Eq. (\ref{DHW}), which are given by 
\begin{align}
    D_t &= \frac{\partial}{\partial t} + e \tilde{\mathbf{E}} \cdot \nabla_p \label{Dt} \\ 
    \tilde{\mathbf{p}} &= \mathbf{p} - ie \int^{1/2}_{-1/2}d\tau \tau \mathbf{B}(\mathbf{r} +i \tau \nabla_p)\times \nabla_p\\
    \textbf{D} &= \nabla_r+ e\int^{1/2}_{-1/2}d\tau \tau \mathbf{B}(\mathbf{r} + i\tau \nabla_p)\times \nabla_p\\
    \tilde{\mathbf{E}} &= \int^{1/2}_{-1/2}d\tau  \mathbf{E}(\mathbf{r} +i \tau \nabla_p). \label{EDtilde}
\end{align}
As before, the operators reduce to their local approximations (i.e. $D_t\rightarrow \partial/\partial t + e \mathbf{E} \cdot\nabla_p$, $\mathbf{D} \rightarrow \nabla_r + e \mathbf{B} \times \nabla_p$, $\tilde{\mathbf{E}} \rightarrow \mathbf{E}$, and $\tilde{\mathbf{p}}\rightarrow \mathbf{p}$) for scale lengths much longer than the characteristic de Broglie length. 

To close the system, we need the source terms in Maxwell's equations, which are given by 
\begin{equation}
\mathbf{j} = \frac{e}{(2\pi)^3} \int d^3p\, \mathbf{v} (\mathbf{p}, \mathbf{r}, t)
\end{equation}
and
\begin{equation}
\label{Conservation_charge}
    \rho = \frac{e}{(2\pi)^3} \int d^3p v_0(\mathbf{p}, \mathbf{r}, t)
\end{equation}
Thus $v_0$ is the time component of the four-vector phase space function that gives the four current density, and ${\bf v}$ is the spatial component, i.e. the current density. Most, if not all, DHW-functions have fairly concrete interpretations which helps guiding the physical intuition. To gain a better understanding, we take a look at some of the conserved quantities of the DHW-system (for a derivation, see Ref. \cite{BB-1991}). Firstly, the total energy $W$ is given by
\begin{equation}
\label{Conservation_Energy}
W = \frac{1}{(2\pi)^3}\int d^3pd^3r \big[ \mathbf{p} \cdot \mathbf{v}(\mathbf{r}, \mathbf{p} ,t) 
+ m s(\mathbf{r}, \mathbf{p}, t)   \big]\\
+ \frac{1}{2} \int d^3r \left[ E^2 + B^2   \right].
\end{equation}
secondly, the linear momentum is 
\begin{equation}
\label{Momentum}
    \mathbf{p}= \frac{1}{(2\pi)^3}\int d^3p d^3r\, \mathbf{p} v_0(\mathbf{r}, \mathbf{p}, t)  + \int d^3r \mathbf{E}\times \mathbf{B}
\end{equation}
and, finally, the total angular momentum $\textbf{M}$ is 
\begin{equation}
\label{Angular_Mom}
    \textbf{M} = \frac{1}{(2\pi)^3}\int d^3pd^3r \Big[\mathbf{r} \times \mathbf{p} v_0(\mathbf{r}, \mathbf{p}, t) + \frac{1}{2}\textbf{a}( \mathbf{r}, \mathbf{p}, t)  \Big]\\ + \int d^3r \, \mathbf{r} \times (\mathbf{E}\times \mathbf{B})
\end{equation}
We can see in Eq. (\ref{Conservation_Energy}) that the current density can be related to the kinetic energy, as expected. However, the role of mass density is played by another function $s$, with no trivial relation to the charge density. As the Dirac field contain both electrons and positrons, the lack of a simple relation between the mass density and the charge density should not be surprising. Nevertheless, in the expression for momentum (\ref{Momentum}), we see that ${\bf p} v_0$ acts as a phase space momentum density. Since electrons and positrons contribute with opposite signs to $v_0$, we realize that electrons and positrons  must have a different dependence on $\bf p$. To be concrete, for electrons and positrons moving in the same direction, we need to shift the momentum dependence for the dependent variables according to $\bf p \rightarrow \bf -p$, as will be illustrated more explicitly below. This is not an issue when solving the DHW-equations, as the DHW-variables generally describe coupled electron and positron states anyway.  However, this insight can be of some importance e.g. when interpreting results, in particular when numerical calculations have been made.  The shift in momentum dependence is consistent with the common interpretation of positrons as being electrons moving backwards in time. 

Another observation that can be made based on the conserved quantities is that the term $\propto (1/2) \bf a$ can be identified as the spin contribution to angular momentum, i.e. $\bf a$ gives the spin density. Finally, the equation for $D_t{\bf t}_1 $ in Eq. (\ref{DHW}) shows a division of the total current density into its free part, magnetization part, and polarization part. To be specific, we can deduce that ${\bf t}_2 $ gives the magnetization, and ${\bf t}_1 $ gives the polarization. For some further discussion of the physical interpretation of the DHW-functions, see Ref. \cite{BB-1991}.

\begin{figure}[tp]
    \begin{center}
        \includegraphics[width=0.45\textwidth]{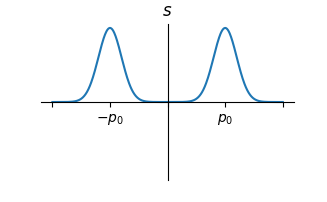}
        \includegraphics[width=0.45\textwidth]{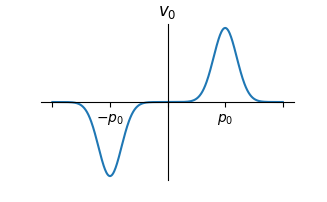}
    \end{center}
    \caption{The distribution functions for the mass density $s$, and the charge density $v_0$ for a beam of electrons and positrons moving in the \textit{same direction} with the common central beam velocity $v = p_0 / \sqrt{m^2 + p_0^2}$.}
    \label{fig:plotsplotv0}
\end{figure}

Contrary to previous kinetic theories, in the DHW-formalism, the kinetic variables are not zero even in vacuum, in older terminology we would say that the vacuum is filled with the particles of the Dirac sea.  However, in the absence of a spin polarizing magnetic field, the only DHW-functions with nonzero vacuum-values are the mass density and current density, which are given by
\begin{align}
s_\text{vac}(\bf p)& =-\frac{2m}{\epsilon }  \notag  \label{Vacuum_sol} \\
\mathbf{v}_\text{vac}(\bf p)& =-\frac{2\mathbf{p}}{\epsilon },
\end{align}%
where $\epsilon=\sqrt{m^2+p^2}$.
The expressions above are obtained by calculating the Wigner operator for the free particle Dirac equation and taking the vacuum expectation value. Note that while the charge density is zero (due to the cancellation of the electron and positron vacuum fluctuations), the same is not true for the current density. The reason is that there are two signs that enter the picture - firstly electrons and positrons have opposite signs of the charge, but secondly, switching electrons for positrons means letting $\bf p \rightarrow \bf -p$, such that the vacuum contributions are additive. Nevertheless, when integrating over momentum to get the total current density, we still get zero as one would expect. 
The substitution $\mathbf{p} \rightarrow - \mathbf{p}$ when switching between electrons and positrons also holds for real particles as well as the vacuum contribution. 
In particular, for beam systems, this is important to keep in mind, as is illustrated in Fig.~\ref{fig:plotsplotv0}.

When adding the real particles of the Dirac field (electrons and/or positrons) into the picture, we can add distribution functions much like in the theories of the previous sections. Specifically, we could start from a background 
\begin{align}
s(\bf p)& =-\frac{2m}{\epsilon }\Big[1-f_{p}({\bf p})-f_{e}(\bf p)\Big]  \notag  \label{Vacuum_sol2} \\
\mathbf{v}(\bf p)& =-\frac{2\mathbf{p}}{\epsilon }\Big[1-f_{p}({\bf p})-f_{e}(\bf p)\Big],
\end{align}
with a nonzero charge density
\begin{equation}
\rho({\bf p}) =2\Big[f_{p}({\bf p})-f_{e}({\bf p})\Big]  
\end{equation}
The function  $f_{e,p}(\bf p)$ can be picked as any common background distribution function from classical kinetic theory, i.e. a Maxwell-Boltzmann, Synge-Juttner, or Fermi-Dirac distribution, depending on whether the characteristic kinetic energy is relativistic and whether the particles are degenerate. 

Note that for a
completely degenerate ($T=0$) Fermi-Dirac background of electrons (and no positrons $f_{p}=0$), the electron and vacuum contributions for the mass density and current density cancel  inside the Fermi sphere. Consequently, for
momenta $p\leq p_{F}$, where  $p_{F}=\hbar (3\pi^2n_{0})^{1/3}$ is the Fermi
momentum, we have $s=\bf v=0$. Furthermore, note that in the presence of a strong field in the background state (e.g. a strong constant magnetic field), these fairly simple background expressions need to be modified. For example, in a case with $\mu_B B_0 \sim mc^2$, also the vacuum states will be subject to Landau quantization. Moreover, the spin density and magnetization (as described by the functions $\bf a$ and $\bf t_1$, respectively)  will no longer vanish in the background state, due to the contribution from real particles. 
 
\subsection{The electrostatic one-dimensional case}
In order to illustrate some features of the DHW-theory, we will consider the case of a one-dimensional electrostatic field, i.e. ${\bf E}=E(z,t){\bf {\hat z}}$. At the same time, the DHW-functions depend on $({\bf p},z,t)$, but where the momentum dependence can be reduced to two independent variables ($p_{\perp},p_z$), due to the rotational symmetry.   

Due to the simplified geometry, only half of the 16 scalar DHW-functions will be nonzero. Moreover, only four of these variables will be independent. There are different ways of finding these nonzero-variables. Here we will just present the result, which is straightforward to verify by direct substitutions into the DHW-equations, Eq. (\ref{DHW}). For a more systematic derivation of the reduced electrostatic equations, see Ref. \cite{Al-Naseri-2021}. As it turns out, the DHW-equations can be expressed in terms of four variables $\chi_1-\chi_4$, related to the original DHW-functions as follows:
\begin{align}
\label{DHW_Non}
    s(z,{\bf p},t)&= \frac{m}{\epsilon_{\bot}}\chi_2(z,{\bf p},t) \notag\\
    v_0(z,{\bf p},t)&= \chi_4(z,{\bf p},t)\notag \\
    \textbf {v}_{\bot}(z,{\bf p},t)&=\frac{ \textbf{p}_{\bot}}{\epsilon_{\bot}}\chi_2(z,{\bf p},t) \notag \\
    v_z(z,{\bf p},t)&=\chi_1(z,{\bf p},t) \\
    a_x(z,{\bf p},t)&=-\frac{p_y}{\epsilon_{\bot}}\chi_3(z,{\bf p},t)\notag\ \\
    a_y(z,{\bf p},t)&=\frac{p_x}{\epsilon_{\bot}}\chi_3(z,{\bf p},t)\notag\\
    t_{1z}(z,{\bf p},t)&=- \frac{m}{\epsilon_{\bot}} \chi_3(z,{\bf p},t) \notag 
\end{align}
where $\epsilon_{\perp}=\sqrt{m^2+p_{\perp}^2}$. Taking this as an ansatz for the variables $\chi_1-\chi_4$ and substituting the expressions into Eq. (\ref{DHW}), we obtain the electrostatic equations: 
\begin{align}
\label{PDE_System}
    D_t\chi_1(z,{\bf p},t)&= 2\epsilon_{\bot}(p_{\bot}) \chi_3(z,{\bf p},t)- \frac{\partial \chi_4}{\partial z} (z,{\bf p},t)\notag\\ 
    D_t\chi_2(z,{\bf p},t) &= -2p_z\chi_3(z,{\bf p},t)\\ 
    D_t\chi_3(z,{\bf p},t)&= -2\epsilon_{\bot}(p_{\bot}) \chi_1(z,{\bf p},t) +2p_z\chi_2(z,{\bf p},t)\notag\\
    D_t\chi_4(z,{\bf p},t)&= -\frac{\partial \chi_1}{\partial z}(z,{\bf p},t)\notag 
\end{align}
where, in the 1D-case of study,  $D_t=\partial_t+eE\partial_{p_z}$
The above system is closed by Ampére's law, which in terms of the $\chi$-variables is written 
\begin{equation}
\label{Ampers_law}
\frac{\partial E}{\partial t}=\frac{e}{(2\pi)^3}\int \chi_1 d^3p
\end{equation}
While Eqs. (\ref{PDE_System})-(\ref{Ampers_law}) represent a huge simplification compared to the full DHW-theory, they can still describe a host of different phenomena. As an example, Ref. \cite{Gies2} has used a similar set to study Schwinger pair-production for given electrostatic  pulses. From a plasma physics perspective, studies of Langmuir waves in a high density plasma is a natural problem. While the nonlinear strong field regime is interesting to study (allowing for the Schwinger mechanism), we leave that for a future numerical investigation. Instead we will focus here on the problem of linearized Langmuir waves, as this is straightforward to study analytically.  

The nonzero background quantities for the $\chi _{i}$-variables can be written as   
\begin{align}
\chi _{1}^{0}(\bf p)& =\frac{2p_{z}}{\epsilon }\Big[f_{p}({\bf p} )+f_{e}({\bf p})-1\Big] 
\notag \\
\chi _{2}^{0}(\bf p) & =\frac{2\epsilon _{\bot }}{\epsilon }\Big[f_{p}({\bf p} )+f_{e}({\bf p})-1\Big]\ \\
\chi_{4}^{0}(\bf p) &=2\Big[f_{p}({\bf p} )-f_{e}({\bf p})\Big]  \notag
\end{align}%
using upper index $0$ for the unperturbed background values, and letting the variables have both electron and positron contributions. 
Next, we divide the variables into unperturbed and perturbed variables
according to 
\begin{equation}
\chi _{i}(z,{\bf p},t)=\chi _{i}^{0}({\bf p})+\chi _{i}^{1}({\bf p})e^{i(kz-\omega t)}
\end{equation}%
(with $\chi _{3}^{0}(\bf p)=0$ and only a perturbed electric field $E$) and linearize
(\ref{PDE_System}) and (\ref{Ampers_law}). Making use of the
relation 
\begin{equation}
\Tilde{\mathbf{E}}\cdot \nabla _{p}\chi _{i}^{0}=\tilde{E}\frac{\partial
\chi _{i}^{0}}{\partial p_{z}}=E\frac{\chi _{i}^{0}(p_{z}+\hbar k/2)-\chi
_{i}^{0}(p_{z}-\hbar k/2)}{\hbar k}  \label{Derivative generalization}
\end{equation}%
the problem is reduced to linear algebra.  Solving for $\chi _{i}^{1}$, and restoring $\hbar$ to identify the quantum contributions (e.g. letting $(2\pi)^3\rightarrow (2\pi\hbar)^3$ in the denominator of Eq. (\ref{Ampers_law})), we
obtain
\begin{widetext}

\begin{align}
    \chi_1&=   \sum_{\pm}
    \frac{\pm i2e\omega E/ (\hbar k) }{(\omega^2-k^2)(\hbar^2\omega^2-4p_z^2)-4\epsilon_{\bot}^2\omega^2}
    \Bigg[ 4p_z\epsilon_{\bot}^2   \frac{F(p_{\pm})}{\epsilon_{\pm}}  
    -(\hbar^2\omega^2 -4p_z^2)
    \bigg( \frac{p_{\pm}}{\epsilon_{\pm}} F(p_{\pm})
  + \frac{k}{\omega}\Big(f_p(p_{\pm})-f_e(p_{\pm})\Big)
    \bigg)
    \Bigg]
    \label{chi1} 
    \\
    \chi_2&=  \sum_{\pm}
    \frac{\mp  i\omega eE \epsilon_{\bot } / (\hbar k) }{(\omega^2-k^2)(\hbar^2\omega^2-4p_z^2)-4\epsilon_{\bot}^2\omega^2}
    \Bigg[  \Big( \hbar^2 \omega^2 -\hbar ^2k^2 - 4\epsilon^2 \mp \frac{\hbar k }{2}p_z\Big)   \frac{F(p_{\pm})}{\epsilon_{\pm}}  
  - 4p_z\frac{k}{\omega}\Big(f_p(p_{\pm})-f_e(p_{\pm})\Big)
    \Bigg]\\
    \chi_3&= \sum_{\pm}
    \frac{\mp  4\omega eE \epsilon_{\bot }  }{(\omega^2-k^2)(\hbar^2\omega^2-4p_z^2)-4\epsilon_{\bot}^2\omega^2}
    \Bigg[  \Big( p_z\frac{k}{\omega}\pm \frac{\hbar \omega }{2}\Big)\frac{F(p_{\pm})}{\epsilon_{\pm}}   + f_p(p_{\pm})-f_e(p_{\pm})
    \Bigg]\\
    \chi_4&= \sum_{\pm}
    \frac{\pm  2i\omega eE/(\hbar k)  }{(\omega^2-k^2)(\hbar^2\omega^2-4p_z^2)-4\epsilon_{\bot}^2\omega^2}
    \Bigg[  \big(4\epsilon^2-\hbar^2\omega^2  \big)
    \left[
    \frac{kp_z}{\omega}\frac{F(p_{\pm})}{\epsilon_{\pm}}  + f_p(p_{\pm})-f_e(p_{\pm})\right] 
    \pm \frac{\hbar k^2}{2\omega}
    \big(4p_z^2-\hbar ^2\omega^2   \big) \frac{F(p_{\pm})}{\epsilon_{\pm}} 
    \Bigg]
    \label{chi4}
\end{align}

where
\begin{align}
p_{\pm}&= p_z\pm \frac{\hbar k}{2}\\
\epsilon_{\pm}&= \sqrt{m^2+p_{\bot}^2 + \Big(p_z \pm \frac{\hbar k }{2}\Big)^2}
\end{align}
Note that $F(p_{\pm})$ and $f_{e,p}(p_{\pm})$ depend on the full momentum, but we suppressed the perpendicular momentum to simplify the notation.
Combining the above results for $\chi_i(\bf p)$ with Ampere's law (\ref{Ampers_law}) we obtain the dispersion relation $D(k,\omega)=0$ with
\begin{multline}
    D(k,\omega)= 1 + \sum_{\pm} \int \frac{d^3p}{(2\pi\hbar)^3}
    \frac{\pm 2e^2 / (\hbar k) }{(\omega^2-k^2)(\hbar^2\omega^2-4p_{\pm}^2)-4\epsilon_{\bot}^2\omega^2}
    \\ 
    \times
    \Bigg[ 4\frac{\epsilon_{\bot}^2}{\epsilon}   p_{\pm} F({\bf p})  
    -(\hbar^2\omega^2 -4p_{\pm}^2)
    \bigg( \frac{p_z}{\epsilon } F({\bf p})
  + \frac{k}{\omega}\Big(f_p({\bf p})-f_e({\bf p})\Big)
    \bigg)
    \Bigg] \label{Full-DR}
\end{multline}
The classical, but relativistic, limit of the dispersion relation is obtained by letting $\hbar \rightarrow 0$. Taking this limit, the dispersion function (\ref{Full-DR}) reduces to
\begin{equation}
    D(k,\omega )= 1+ \frac{e^2}{\omega } \int \frac{d^3p}{(2\pi\hbar)^3} \frac{p_z}{\epsilon}
    \bigg(\frac{1}{\omega-kp_z/\epsilon}+ \frac{1}{\omega+kp_z/\epsilon}
    \bigg)
    \bigg[ \Big(1 + \frac{kp_z}{\epsilon \omega}
    \Big)\frac{\partial f_p({\bf p})}{\partial p_z}
    + \Big(1 - \frac{kp_z}{\epsilon \omega}
    \Big)\frac{\partial f_e({\bf p})}{\partial p_z}
    \bigg], 
\end{equation}
\end{widetext}
which can be shown to agree with the standard result after some straightforward algebra. Note that the appearance of $\hbar$ in the integration measure $\frac{d^3p}{(2\pi\hbar)^3}$ is just a matter of normalization, and not a sign of any remaining quantum features.  

The main purpose here as been to demonstrate the usefulness of the DHW-equation for practical plasma calculations. However, before ending the discussion, without going into details, let us point out a few features of the general dispersion relation. 

\begin{enumerate}
    \item Apart from the effect of a relativistic Fermi velocity, in the relativistic regime, the quantum contribution to the {\it real part of the plasma frequency} will actually decrease with increasing density, since quantum terms are compared with the (high) relativistic Fermi energies and Fermi momenta.     
    \item The main new effect due to the quantum relativistic regime, comes from the new types of resonant denominators, associated with wave damping. In particular, even for $k=0$ we may still have a resonant denominator (corresponding to electron-positron pair-production), provided the pair-creation condition $\hbar \omega\geq 4mc^2$ is fulfilled.   
    \item Due to the vacuum background, the integrand is nonzero even in the absence of particles, giving raise to the effect of vacuum polarization contribution. While this term typically gives a contribution that is much smaller than that from the real particles,  the given expression is subject to ultra-violet divergences that must be handled using a renormalization scheme, see e.g. Ref. \cite{BB-1991}.   
    
\end{enumerate} 

Naturally, the above points only give the principal features. A more thorough study of the quantum relativistic dispersion relation for Langmuir waves has to be done numerically. This will be the subject of a future paper. 

\subsection{Limiting cases of the DHW-theory}

Apart from exchange effects, the DHW-equations covers all the physical phenomena presented in previous sections. Thus, ideally, we should be able to recover all the previous models, (except the parts presented in section III)  as special cases of the DHW-formalism. However, demonstrating the equivalence in
appropriate limiting cases is somewhat non-trivial. First of all, the
DHW-equations do not only describe electrons, since a Foldy-Wouythuysen
transformation cannot be made in the fully quantum relativistic regime. Furthermore, the equations with a spin-dependent Wigner-function
uses a Q-transform to get a scalar theory, which further complicates a
comparison. Nevertheless, though a complete investigation is yet to be made,
showing the equivalence of the DHW-formalism with the models of section IV (i.e. the non-relativistic Pauli-limit) is relatively straightforward.

First, we note that for fields well below the critical field, there is 
little ambiguity whether we have electron or positron states.
Considering the case of electrons only, ignoring relativistic effects, the charge and mass density are the same (due to the normalization, the constant factors involving $e$ and $m$ do not enter), i.e. $s=v_0$. As a first exercise, let us recover the model based on the Schrödinger Hamiltonian. This implies
dropping the effects of the electron spin, i.e. the spin density, magnetization, and spin polarization are zero, and thus we let ${\bf a}={\bf t_2}={\bf t_1}=0$ in the DHW-equations.  As a result of the above approximations, we get  ${\bf v}={\bf {\tilde p}}s={\bf {\tilde p}}v_0$, which immediately lead to a closed equation for $v_0$:  
\begin{equation}
D_t v_0+{\bf D}\cdot({\bf {\tilde p}}v_0)=0
\label{Limit1}
\end{equation}
Identifying $v_0$ with the Wigner function of section II, using the definitions of the nonlocal variables to write the more explicitly, Eq. (\ref{Limit1}) gives us
\begin{equation}
	\frac{\partial f}{\partial t} + (\mathbf v + \Delta \tilde{\mathbf v} ) \cdot \nabla_x f + 
	\frac{q}{m} \left[ (\mathbf v + \Delta \tilde{\mathbf v} ) \times \tilde{\mathbf B} + 
	\tilde{\mathbf E} \right] \cdot \nabla_v f =0
	\label{Limit2}
\end{equation}
The definitions of $\Delta {\bf {\tilde v}}$, etc., are the same as those in section II. We note that Eq. (\ref{Limit2}) coincides with Eq, (\ref{gaugeevolu}). Thus Eq. (\ref{Limit2}) generalizes the results for the electrostatic version of the Schrödinger Hamiltonian to also cover electromagnetic fields, as follows naturally using the gauge-invariant Wigner transform, as discussed in section \ref{IIC}. Alternatively, we could re-derive the same equation by dropping the spin terms in Eq. (\ref{82}) (all terms proportional to $\mu_B$), and integrate the equation over spin space.

Next, our aim is to recover the model of Section IV, based on the Pauli Hamiltonian. As
the short scale physics associated with the non-local expressions (\ref{Dt})-(\ref{EDtilde})
has already been established above, for convenience we restrict ourselves to the
case of long scale lengths \ ($\hbar \nabla \nabla_p \ll 1$), such that the local approximations (i.e. dropping the variables with tilde) can be
used. For the non-relativistic case, we can still put the charge and mass density equal, i.e. $s=v_0$. Secondly, in the absence of relativistic effects (and for the given normalizations) and for the case of electrons only, the spin density and  the magnetization is the same, i.e. ${\bf a}={\bf t}_2$.
Thirdly, a non-vanishing polarization due to the spin only enters in the relativistic theory, and thus we can put ${\bf t}_1=0$. Finally,  the term $\propto D_t  \rho $ in Eq. (\ref{DHW}) is a small correction (in a quantum relativistic expansion $\hbar \partial_t/mc^2$), and hence we can use the approximation $a_0={\bf p}\cdot{\bf t}_1/m$. With these simplifications as a starting point, the DHW-equations can be combined to give the following evolution equation for the magnetization: 
\begin{equation}
D_t{\bf t}_2 +{\bf D}({\bf p}\cdot{\bf t}_2)+{\bf p}\times\bigg[{\bf p}\frac{2v_0}{m}+\frac{{\bf D}\times{\bf t}_2}{m}\bigg]=0 \label{DHW-magn}
\end{equation}
Normally, we should drop the term $\propto {\bf p}\times {\bf p}$, since such a term is identically zero. However, recall that before introducing approximations, this term would rather be proportional to  ${\bf {\tilde p}}\times {\bf {\tilde p}}$. As it turns out, the local approximations are applicable everywhere else, as in those cases, the corrections are compared with larger surviving terms. Here, however, we need to use the full expression $\propto {\bf {\tilde p}}\times {\bf {\tilde p}}$, and evaluate the term to the first non-vanishing order in an expansion in the small parameter $\hbar\nabla\nabla_p$.  Performing this expansion and identifying ${\bf t}_2$ with ${\bf f}$ and $v_0$ with $f_0$, after some algebra we can confirm the exact agreement of Eq. (\ref{DHW-magn}) with Eq. (\ref{Manfred-vector}). 

Next, to establish agreement with the model based on the Pauli Hamiltonian, we need to re-derive Eq. (\ref{Manfred-scalar}). Using the same approximations as before, we immediately get
\begin{equation}
D_t v_0+{\bf D}\cdot({\bf p}v_0)
+{\bf D}\cdot({\bf D}\times{\bf t}_2)=0
\label{DHW-scalar}
\end{equation}
While there is a litle bit of algbra involved (since the operator ${\bf D}$ contains the magnetic field), it is straightforward to show that Eq. (\ref{DHW-scalar}) reduces to Eq. (\ref{Manfred-scalar}). Finally, we note that the current sources to be used in the DHW-equations are the same as in the Pauli-limit, given that polarizations currents are dropped in non-relativistic theory. Furthermore, as the agreement between Eqs. (\ref{Manfred-scalar})-(\ref{Manfred-vector}) and (\ref{paulilong}) has already been established, and the short-scale physics (as described by Eq. (\ref{Limit2})) have been recovered, for most practical purposes, the full model based on the Pauli Hamiltonian (Eq. (\ref{eq:Ham-Pauli})) have also been verified, although not in a strict sense. 

While the above confirmation is reassuring, the main purpose of studying  approximate versions of the DHW-model is to find models that are easier to analyze in specific cases. The DHW-equations allow for systematic expansions in numerous quantum and relativistic parameters, $\hbar\omega/mc^2$, $\hbar e{F}/mc$ (where $F$ represents electric and/or magnetic fields),  $\hbar\nabla\nabla_p$, $p/mc$, and $\hbar k^2/m\omega$ to name a few. Here $\omega$ and $k$ represents general temporal and spatial scales, rather than specific frequencies and wave numbers. Depending on the ordering of the dimensionless parameters, there are several possibilities for approximate models containing different combinations of expansion parameters. A systematic search for consistent approximations of the DHW-system is a project for further research.

\section{Concluding remarks}
\label{section7}

While the aim of this review has been to cover many aspects of quantum kinetic theory, naturally, there are many interesting and important topics that we have not touched upon. While, undoubtedly, many things will be left out altogether, let us here at least partially remedy some of the omissions mad, by pointing to a few relevant aspects of quantum kinetic theory that we have not covered. 

Firstly, as is well-known, there are close connections between hydrodynamic and kinetic theories. In particular, a common approach to derive accurate fluid theories is by making moment expansions of kinetic theories. In a quantum context, this scheme has been used to derive fluid theories from kinetic theories e.g. by Refs. \cite{Momexp1,Momexp2,Momexp3} for the model defined by the Schrödinger Hamiltonian and by Refs.  \cite{Manfredi-spin1,Momexp4} for the Pauli Hamiltonian model, and also by Ref. \cite{Hurst-2017} for the model where the Pauli-Hamiltonian is extended by the spin-orbit term. Moreover, moment expansion for models including exchange effects has been made by e.g. Refs. \cite{Haas-exchange,Manfredi_DFT}.

Secondly, we note that for completely degenerate systems, in certain cases kinetic models may be simplified. This happens in situations when the phase-space density is conserved, in which case the dynamics is determined by the Fermi surface. This has been explored in so called waterbag models of plasmas \cite{Manfredi_DFT,Waterbag} and also for the case of a semiclassical model based on the Pauli Hamiltonian \cite{Brodin-Stefan}. 

Thirdly, we stress that we here have focused on non-dissipative quantum kinetic models, ignoring the effects of higher order correlation in the BBGKY-hierarchy, neglecting all the influence of collisions. The effect of dissipation is a research topic in its own right, of particular importance in the strong coupling regime. For references covering this field, see e.g. \cite{Bonitz-book,Ichimaru82,Bonitz2015}.

As our review has focused on quantum kinetic models, and the worked out examples has served the purpose of illustrating the applicability of the theory, many aspects of quantum kinetic theory have still been ignored.  In particular interesting theoretical aspects involving wave-particle interaction, spin-polarization dynamics, numerical computation schemes, and the interesting interplay with single particle dynamcis \cite{QRR}

\appendix
\section{BBGKY-Hierarchy and the Mean-Field Approximation} 
\label{appendix}
The $N$-particle density matrix is not directly applicable in the case where we have a large number of particles, as is typically the case for plasmas. 
It is therefore necessary to make some approximations to reduce the complexity of the equations. 
One commonly used approximations when dealing with plasmas is the mean-field approximation, which we will consider here. 
Since it is more convenient for this type of considerations, we will here use the operator representation of the density matrix (as opposed to using the position representation as was done in Section~\ref{section2}.
Using the bra-ket notation we can define the \textit{density operator} as 
\begin{equation}
    \hat{\rho} = \sum_i p_i \left| \psi_i \right> \left< \psi_i \right|, 
\end{equation}
where $\left| \psi_i \right>$ are $N$-particle states. 
The evolution equation can then be written as
\begin{equation}
    i \hbar \partial_t \hat{\rho} = \left[ \hat{\rho}, \hat{H} \right] , 
    \label{vonneumann1}
\end{equation}
where $\left[ \cdot , \cdot \right]$ denotes the commutator. 
Assume that we have an Hamiltonian in the form
\begin{equation}
    \hat H_{1 \dots N} = \sum_{i = 1}^N \hat{H}_i + \frac{1}{2} \sum_{\substack{i,j=1 \\ i \neq j}}^N \hat{V}_{ij} ,   
\end{equation}
where 
\begin{equation}
    \hat{H}_i = \frac{\hat{\mathbf{p}}^2_i}{2m}, 
\end{equation}
is the kinetic energy operator, with $\hat{\mathbf{p}}_i$ being the momentum operator acting on the $i$:th particle. 
The operator in the last term $\hat{V}_{ij}$ corresponds to particle-particle interactions, which we will later take as the Coulomb interaction. 
However, for now, we only need to assume that the interaction is symmetric under exchange of particles, i.e., $\hat{V}_{ij} = \hat{V}_{ji}$. 
We note that the density matrix is symmetric with respect to change of particles, i.e., 
\begin{equation}
	\hat{\rho}_{1, \dots i, \dots j, \dots N} = \hat{\rho}_{1, \dots j, \dots, i, \dots N}. 
\end{equation}
This is true, as long as the particles are indistinguishable, and hence both for \textit{fermions} and \textit{bosons}. 
We may now define the $s$-particle \textit{reduced density matrix} as
\begin{equation}
    \hat{\rho}_{1\dots s} = N^i \mathrm{Tr}_{s + 1 \dots N} \hat{\rho}_{1\dots N}, 
    \label{reduced1} 
\end{equation}
where we, for convenience, have multiplied by the number of particles $i$ times. 
Now, using the above we can derive the so called BBGKY-hierarchy for the reduced density matrices: 
\begin{equation}
	i \hbar \partial_t \hat{\rho}_{1 \dots s} = \sum_{i=1}^s \left[ \hat{H}_i , \hat{\rho}_{1 \dots s} \right] 
	+ \mathrm{Tr}_{s + 1} \sum_{i=1}^s \left[ \hat{V}_{i, s+1} , \hat{\rho}_{1 \dots s+1} \right] . 
\end{equation}
Here we have assumed that $N$ is large so that we may use $N-s \approx N$ to any order $s$ where the above equation is 
of any practial use. 
We note that the one-particle density matrix couples to the two-particle density matrix, etc. 
Now, we are only interested in the lowest equation which is
\begin{equation}
	i \hbar \partial_t \hat{\rho}_{1} = \left[ \hat{H} , \hat{\rho}_{1} \right] 
	+ \mathrm{Tr}_{2} \left[ V_{12} , \rho_{12} \right] . 
\end{equation}
In order to make progress from this, we write the two-particle density matrix as
\begin{equation}
	\hat{\rho}_{12} = \hat{\rho}_1 \hat{\rho}_2 + \hat{g}_{12} , 
\end{equation}
where $\hat{g}_{12}$ is the two-particle correlations defined by the equation above. 
In the lowest order approximation, we completely neglect the two-particle correlations $\hat{g}_{12} \approx 0$. 
We then get the so-called mean-field, or Hartree approximation
\begin{equation}
	i \hbar \partial_t \hat{\rho}_1 = \left[ \frac{\hat{\mathbf{p}}^2}{2m}, \hat{\rho}_1 \right] 
	+ \hat{\rho}_1 \mathrm{Tr}_2 \left[ V_{12}, \hat{\rho}_2 \right].
\end{equation}
We note that the normalisation of the density matrix is such that 
\begin{equation}
	\mathrm{Tr}_1 \hat{\rho}_1 = N,
\end{equation}
i.e., the number of particles. 
The diagonal elements can then be identified as the local density of particles $\rho(\mathbf{r}, \mathbf{r}) = n(\mathbf{r})$. 
In particular, in the case of a plasma interacting via the Coulumb interaction 
\begin{equation}
	\hat{V}_{12} = \frac{e^2}{4 \pi \epsilon_0 \left| \hat{\mathbf{r}}_1 - \hat{\mathbf{r}}_2 \right|}, 
\end{equation}
and the last term is
\begin{equation}
	\rho(\mathbf{r}_1 , \mathbf{r}'_1) 
	e \int d^3 r_2 \frac{e n(\mathbf{r}_2)}{4 \pi \epsilon_0} \left( \frac{1}{\left| \mathbf{r}_1 - \mathbf{r}_2 \right|}
	- \frac{1}{\left| \mathbf{r}'_1 - \mathbf{r}_2 \right|}
	\right)
	= \rho(\mathbf{r}_1, \mathbf{r}'_1) e \left[ V_{mf} (\mathbf{r}_1) - V_{mf}  (\mathbf{r}'_1) \right],
\end{equation}
where the last step is written in terms of the mean-field potential, i.e., the potential created by all particles interacting with particle 1.  
Here we have only considered particles which are indistinguishable, but we have not yet taken into account the correct symmetry. 
For fermions, we need to modify the procedure slightly in order to account for the antisymmetry of the wave function, but the basic principle is the same. 
We will do this modification in Section~\ref{section3} where we consider exchange effects. The conclusion from the above considerations is that if we can assume that particle correlations are negligible, we may describe the plasma as particles moving in the mean-field created by all the other particles.

\end{document}